\documentclass[aps,twocolumn,pra,showpacs,superscriptaddress,10pt]{revtex4-1}
\usepackage{amssymb,mathrsfs}

\usepackage{amsmath}
\usepackage{graphicx,color}
\usepackage{epstopdf}
\usepackage{mathrsfs}
\usepackage{epsfig}
\usepackage[breaklinks=true]{hyperref}

\newcommand{\beq}{\begin{equation}}
\newcommand{\eeq}{\end{equation}}
\newcommand{\bqa}{\begin{eqnarray}}
\newcommand{\eqa}{\end{eqnarray}}

\definecolor{green}{rgb}{0.00,0.50,0.00}

\begin{document}

\title{Feedback Network Models for Quantum Transport}
\date{\today}
\author{John Gough} \email{jug@aber.ac.uk}
\affiliation{
  Aberystwyth University, 
	Aberystwyth,
  SY23 3BZ,
  United Kingdom
}
\date{\today}
\begin{abstract}
Quantum feedback networks have been introduced in quantum optics as a set of rules for constructing arbitrary networks
of quantum mechanical systems connected by uni-directional quantum optical fields, and has allowed for a system theoretic
approach to open quantum optics systems. Our aim here is to establish a network theory for quantum transport systems where
typically the mediating fields between systems are bi-directional. Mathematically this leads us to study quantum feedback networks where 
fields arrive at ports in input-output pairs, which is then just a specially case of the uni-directional theory. However, it is conceptually 
important to develop this theory in the context of quantum transport theory, and the resulting theory extends traditional approaches
which tends to view the components in quantum transport as scatterers for the various fields, in the process allows us to consider 
emission and absorption of field quanta by these components. The quantum feedback network theory is applicable to both Bose and Fermi
fields, moreover it applies to nonlinear dynamics for the component systems. In this first paper on the subject, we advance the general theory, but 
study the case of linear passive quantum components in some detail.
\end{abstract}
\maketitle

\section{Introduction}
The aim of this paper is to extend the formalism of quantum feedback networks  \cite{GJ-QFN},\cite{GJ-Series} 
from their current applications to quantum optical, and more recently
opto-mechanical systems, to quantum transport networks. In the quantum optics
applications, one usually treats the noise fields interacting with the system
as uni-directional. In the input-output approach of Gardiner and Collett, see \cite{GarZol00},
this arises naturally and may be understood as a specific case of the LSZ
formalism of quantum field theory, however, physically this is also justified
by the fact that bi-directional quantum optical fields may be always made
uni-directional by using an optical isolator.

The quantum feedback network theory is built on the general theory of
open quantum stochastic evolutions developed by Hudson and Parthasarathy \cite{HP}
which goes beyond Gardiner's theory by allowing the system to scatter noise
quanta as well as emit and absorb them - now generally referred to as the SLH
formalism, which we recall in the next section.

There has been increasing interest in developing control theory for quantum transport 
models. The control of solid state cavity QED devices in place of traditional photonic systems, see e.g. \cite{Wetal}
for super-conducting qubit examples,
has started to lead to some of the techniques applied to control quantum optical devices 
being applied in new settings. Coupling a QED cavity to a quantum dot
has been shown to allow control of the cavity reflectivity \cite{EFF}, as well as the possibility to generate
non-classical states of light \cite{Fetal}, see \cite{OFV} and \cite{V} for an overview of recent applications to
photonics and quantum dots in photonic crystal technologies. Quantum dots have also been used to stabilise mesoscopic electric currents 
by means of feedback \cite{Brandes}, with proposals for delayed feedback \cite{Emary} and stabilisation of 
pure qubit states \cite{PCB}. As with quantum optical devices, there has been a move away from table top experimental set-ups
towards on-chip devices, and strong photon–photon interactions have been shown to be implementable on integrated photonic chips were
quantum dots embedded in photonic-crystal nanocavities \cite{Retal}.

A first step in extending quantum feedback 
networks to quantum transport problems has been made in \cite{EG}, which treated
the scattering of the noise quanta only, but considered control methodologies.
Here, however, we wish to extend the theory to general linear systems which 
allow for more general models of dissipation. This leads to the framework in which to apply the standard techniques of measurement-based
and coherent quantum feedback techniques. We expect that the theory presented here should be readily implementable
with existing toolboxes for simulating quantum feedback networks \cite{Tezak}, \cite{Tezak1}.

Although the theory is applicable to general coupling of the fields to the components, we will develop the linear
theory in some detail. Here the chain scattering representation proves to be essential approach. We point out 
that there exists a well-developed theory of control based on this approach due to Kimura \cite{Kimura}, and which
we exploit here. The results on lossless systems is particular relevant to the linear passive models which we consider here.
We also wish to acknowledge the prior work of Yanagisawa and Kimura \cite{YK1}, \cite{YK2}
on quantum linear models which as far as we know was the first
to apply chain scattering techniques to linear quantum networks.

For transparency we restrictive to passive systems \cite{GGY}, however it it clear that many of the results presented here extend 
should carry over to quantum transport networks which include active components \cite{GJN10}.

\section{The SLH-Formalism}

For open Markov systems driven by $n$ vacuum noise inputs, the model is
specified by a triple 
\begin{equation*}
\mathbf{G}\sim (S,L,H) 
\end{equation*}
referred to as the set of Hudson-Parthasarathy coefficients, or more prosaically
as the ``SLH''. Their roles are to describe the
input-to-output scattering $S= [S_{jk}]$ of the external noise fields 
$b_{k}\left( t\right) $, the coupling $L= [L_j ]$ of the noise to the system,
and the internal Hamiltonian $H$ of the system respectively.

The SLH formalism for quantum Markov models deals with the category of models

\begin{equation*}
S=\left[ 
\begin{array}{ccc}
S_{11} & \cdots  & S_{1n} \\ 
\vdots  & \ddots  & \vdots  \\ 
S_{n1} & \cdots  & S_{nn}
\end{array}
\right] ,L=\left[ 
\begin{array}{c}
L_{1} \\ 
\vdots  \\ 
L_{n}
\end{array}
\right] ,H ,
\end{equation*}
where the $S_{jk},L_k,H$ are operators on the component system Hilbert space.

\begin{figure}[h]
	\centering
		\includegraphics[width=0.30\textwidth]{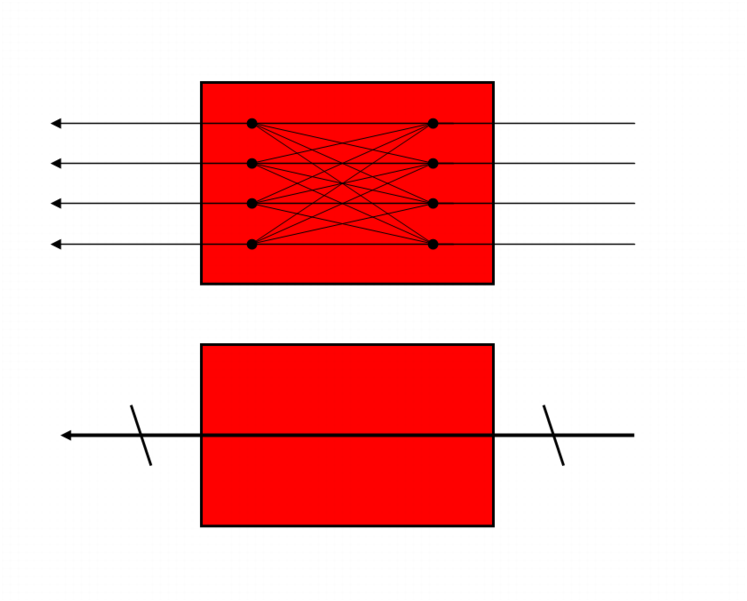}
		\caption{(color online) A component representing a quantum mechanical system driven by
		several input fields. There will be the same number of output fields. It is often convenient to think of grouped
		inputs with multiplicity greater than one.}
	\label{fig:QFN_multi}
\end{figure}

These may be assimilated into the \textit{model matrix}
\begin{widetext}
\begin{eqnarray*}
\mathsf{V} =\left[ 
\begin{array}{cc}
-\frac{1}{2}L^{\ast }L-iH & -L^{\ast }S \\ 
L & S
\end{array}
\right] = \left[ 
\begin{array}{cccc}
-\frac{1}{2}\sum_{j}L_{j}^{\ast }L_{j}-iH & -\sum_{j}L_{j}^{\ast }S_{j1} & 
\cdots  & -\sum_{j}L_{j}^{\ast }S_{jm} \\ 
L_{1} & S_{11} & \cdots  & S_{1n} \\ 
\vdots  & \vdots  & \ddots  & \vdots  \\ 
L_{n} & S_{n1} & \cdots  & S_{nn}
\end{array}
\right]  =\left[ 
\begin{array}{cccc}
\mathsf{V}_{00} & \mathsf{V}_{01} & \cdots  & \mathsf{V}_{0m} \\ 
\mathsf{V}_{10} & \mathsf{V}_{11} & \cdots  & \mathsf{V}_{1n} \\ 
\vdots  & \vdots  & \ddots  & \vdots  \\ 
\mathsf{V}_{n0} & \mathsf{V}_{n1} & \cdots  & \mathsf{V}_{nn}
\end{array}
\right] .
\end{eqnarray*}
\end{widetext}

We recall briefly the class of Markov models for open quantum systems. The
system with Hilbert space $\mathfrak{h}$ driven by $n$ independent Bose
quantum processes with Fock space $\mathfrak{F}$ will have a unitary
evolution $V_{\mathbf{G}}(t)$\ on the space $\mathfrak{h}\otimes \mathfrak{F}
$ where $V_{\mathbf{G}}(t)$ is the solution to the quantum stochastic
differential equation \cite{HP}

\begin{gather*}
dV_{\mathbf{G}}(t)=\{ (S_{jk}-\delta _{jk})\otimes d\Lambda
_{jk}(t)+L_{j}\otimes dB_{j}^{\ast }(t) \\
 -L_{j}^{\ast }S_{jk}\otimes dB_{k}(t)-(\frac{1}{2}L_{k}^{\ast
}L_{k}+iH)\otimes dt \} \, V_{\mathbf{G}}(t)
\end{gather*}

with initial condition $V_{\mathbf{G}}(0)=I$. (We adopt the convention that
repeated Latin indices imply a summation over the range $1,\cdots ,n$.)
Formally, the Bose noise can be thought of as arising from quantum white
noise processes $b_{k}(t)$ satisfying a set singular of commutation
relations 
\begin{equation*}
\left[ b_{j}(t),b_{k}(s)^{\ast }\right] =\delta _{jk}\delta (t-s), 
\end{equation*}
with 
\begin{eqnarray*}
B_{j}(t) &=&\int_{0}^{t}b_{j}(s)ds,\quad B_{j}^{\ast
}(t)=\int_{0}^{t}b_{j}(s)^{\ast }ds, \\
\Lambda _{jk}(t) &=&\int_{0}^{t}b_{j}(s)^{\ast }b_{k}(s)ds.
\end{eqnarray*}
The conditions guaranteeing unitarity are that $S=\left[ S_{jk}\right] $ is
unitary, $L=\left[ L_{j}\right] $ is bounded and $H$ self-adjoint. In the
autonomous case we may assume that the operator coefficients $S_{jk},L_{j},H$
are fixed system operators, however there is little difficulty in allowing
them to be time dependent, or more generally be adapted processes, that is 
$S_{jk}(t),L_{j}(t),H(t)$\ depend on the noise up to time $t$. The process 
$V_{\mathbf{G}}(t)$ will inherit this adaptedness property.

For a fixed system operator $X$ we set 
\begin{equation}
j_{t}^{\mathbf{G}}(X)\triangleq V_{\mathbf{G}}(t)^{\ast }
\left[ X\otimes I\right] V_{\mathbf{G}}(t).
\end{equation}
Then from the quantum It\={o} calculus \cite{HP} we get the \textit{%
Heisenberg-Langevin Equations} 
\begin{eqnarray}
dj_{t}^{\mathbf{G}}(X) &=&j_{t}^{\mathbf{G}}(\mathcal{L}_{jk}X)\otimes
d\Lambda _{jk}(t)+j_{t}^{\mathbf{G}}(\mathcal{L}_{j0}X)\otimes dB_{j}^{\ast
}(t)  \notag \\
&&+j_{t}^{\mathbf{G}}(\mathcal{L}_{0k}X)\otimes dB_{k}(t)+j_{t}^{\mathbf{G}}
(\mathcal{L}_{00}X)\otimes dt  \label{dynamical}
\end{eqnarray}
where the Evans-Hudson superoperators $\mathcal{L}_{\mu \nu }$ are
explicitly given by 
\begin{eqnarray*}
\mathcal{L}_{jk}X &=&S_{lj}^{\ast }XS_{lk}-\delta _{jk}X, \\
\mathcal{L}_{j0}X &=&S_{lj}^{\ast }[X,L_{l}], \\
\mathcal{L}_{0k}X &=&[L_{l}^{\ast },X]S_{lk} \\
\mathcal{L}_{00}X &=&\frac{1}{2}L_{l}^{\ast }[X,L_{l}]
+\frac{1}{2}[L_{l}^{\ast },X]L_{l}+i\left[ X,H\right] .
\end{eqnarray*}
In particular $\mathcal{L}_{00}$ takes the generic form of a Lindblad
generator.

The \emph{output processes} are then defined to be 
\begin{equation*}
B_{j}^{\mathrm{out}}(t) \triangleq V_{\mathbf{G}}(t) ^{\ast }
\left[ I\otimes B_{j}(t) 
\right] V_{\mathbf{G}}(t) , 
\end{equation*}
Again using the quantum It\={o} rules, we see that 
\begin{equation*}
dB_{k}^{\mathrm{out}}=j_{t}^{\mathbf{G}}(S_{kl})dB_{l}(t)
+j_{t}^{\mathbf{G} }(L_{k})dt. 
\end{equation*}

The input-output relations for the column vector $B^{\mathrm{out}}=\left[
B_{j}^{\mathrm{out}}\right] $can be written as a \textit{Galilean
transformation} 
\begin{equation*}
\left[ 
\begin{array}{c}
dt^{\mathrm{out}} \\ 
dB^{\mathrm{out}}(t)
\end{array}
\right] =j_{t}^{\mathbf{G}}(M)\left[ 
\begin{array}{c}
dt \\ 
dB(t)
\end{array}
\right] ,\quad M=\left[ 
\begin{array}{cc}
1 & 0 \\ 
L & S
\end{array}
\right] . 
\end{equation*}

\subsection{Networks}
The rules for construction arbitrary network architectures  were derived in \cite{GJ-QFN}.

\subsubsection{The Parallel Sum Rule}
\begin{figure}[h]
	\centering
		\includegraphics[width=0.30\textwidth]{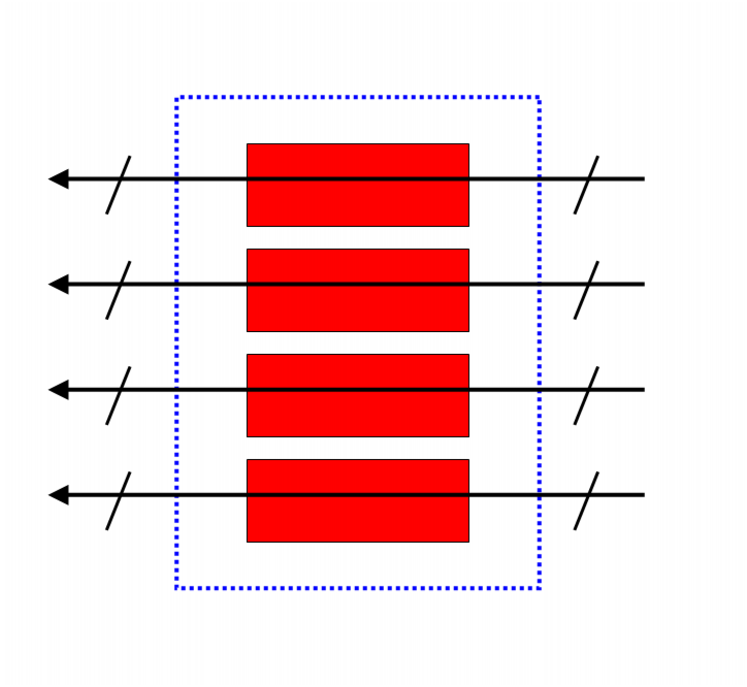}
		\caption{(color online) Several SLH models run in parallel: they correspond to one single SLH model.}
	\label{fig:QFN_Parallel}
\end{figure}

If we have several quantum Markov models with independent inputs then they may be assembled into
a single SLH model, see Fig. \ref{fig:QFN_Parallel}.

\begin{eqnarray*}
&& \boxplus _{j=1}^{n}\left( S_{j},L_{j},H_{j}\right) =\\
&& \left( \left[ 
\begin{array}{ccc}
S_{1} & 0 & 0 \\ 
0 & \ddots  & 0 \\ 
0 & 0 & S_{n}
\end{array}
\right] ,\left[ 
\begin{array}{c}
L_{1} \\ 
\vdots  \\ 
L_{n}
\end{array}
\right] ,H_{1}+\cdots +H_{n}\right) .
\end{eqnarray*}

Note that the components need not be distinct - that is, observable associated with one component are not assumed to commute with those of others.
In this case the definition is not quite so trivial as it may first appear.

\subsubsection{The Feedback Reduction Rule}
\begin{figure}[h]
	\centering
  	\includegraphics[width=0.40\textwidth]{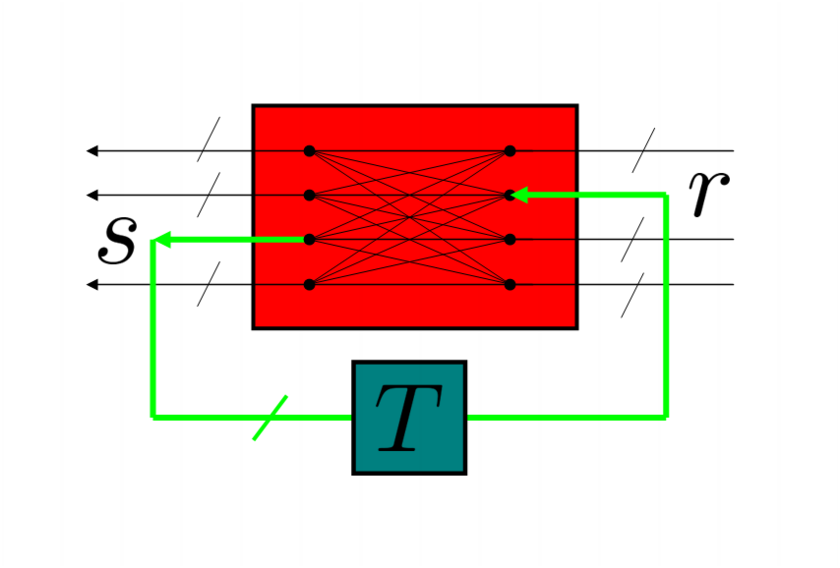}
  	\caption{(color online) We feed selected outputs back in as inputs to get a reduced model.}
	\label{fig:QFN_FR}
\end{figure}

If we wish to feedback an output back in as an input, we obtain a reduced model as depicted in Fig. \ref{fig:QFN_FR}.

The feedback reduction yields the model matrix \cite{GJ-QFN}
\begin{equation}
\left[ \mathscr{F}_{\left( r,s\right) }(\mathsf{V},T)\right] _{\alpha
\beta }=\mathsf{V}_{\alpha \beta }+\mathsf{V}_{\alpha r}T\left( 1-\mathsf{V}%
_{rs}T\right) ^{-1}\mathsf{V}_{s\beta }
\label{eq:QFN_FR}
\end{equation}
for $\alpha \neq r$ and $\beta \neq s$. We remain in the category of SLH models provided that $T$ is unitary and
the network is well-posed, that is $1-\mathsf{V}_{rs}T$ is invertible.

\subsubsection{Construction}
\begin{figure}[h]
	\centering
		\includegraphics[width=0.40\textwidth]{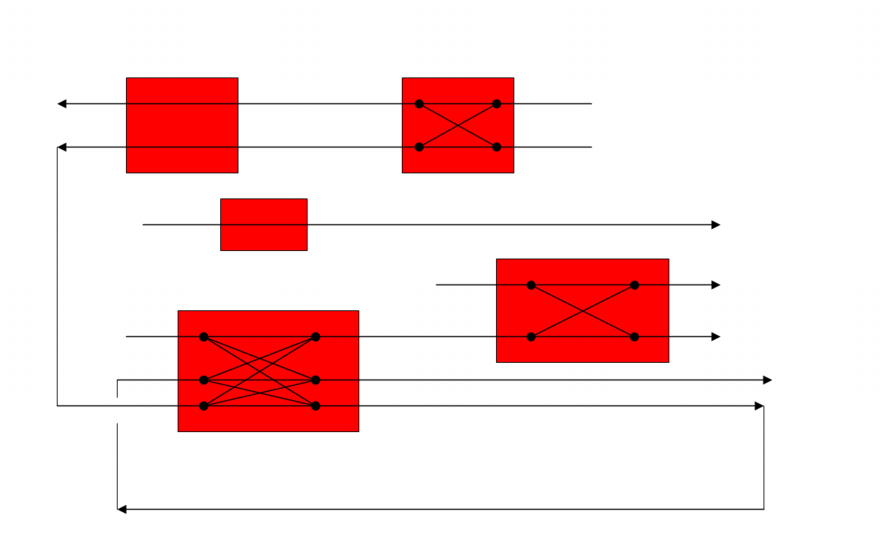}
	\caption{(color online) An arbitrary quantum feedback network.}
	\label{fig:QTFN_QFN_network}
\end{figure}

If, for instance, we wished to determine the effective SLH model for the network shown in Fig. \ref{fig:QTFN_QFN_network},
then we would proceed as follows: first of all we disconnect all the internal lines, this leaves us with an ``open-loop'' description
where all the components are have independent inputs and outputs, and at this stage we use the parallel sum to collect all these components into
a single open-loop quantum Markov component; the next step is to make the connections and this involves feeding selected outputs back in as inputs
from the open-loop description, and to this end we use the feedback reduction formula. This process has recently been automated using a workflow
capture software QHDL \cite{Tezak} and \cite{Tezak1}.

\subsection{Systems in Series}

\begin{figure}[htbp]
	\centering
		\includegraphics[width=0.40\textwidth]{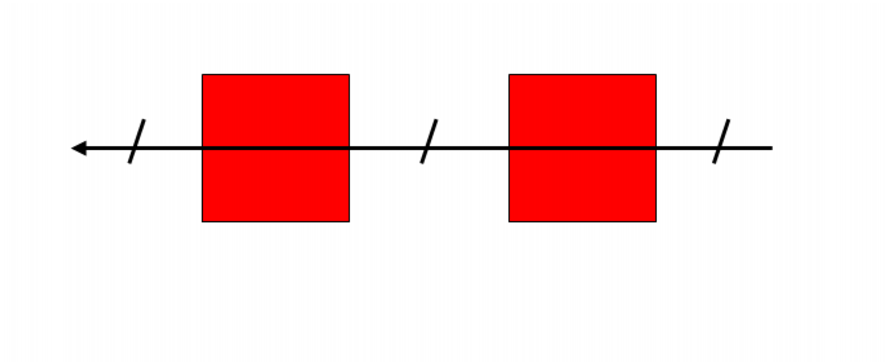}
	\caption{(color online) Systems in series.}
	\label{fig:QTFN_QFN_series}
\end{figure}

The simplest model consists of two systems cascaded together as shown in Fig. \ref{fig:QTFN_QFN_series} and is equivalent to the single component, 
see \cite{GJ-Series},
\begin{multline*}
\left( S_{2},L_{2},H_{2}\right) \vartriangleleft \left(
S_{1},L_{1},H_{1}\right) = \\
\left( S_{2}S_{1},L_{2}+S_{2}L_{1},H_{1}+H_{2}+\text{Im}\left\{ L_{2}^{\dag
}S_{2}L_{1}\right\} \right) .
\end{multline*}
We refer to $G = G_2 \vartriangleleft G_1$ above as the series product of the $G_1$ and $G_2$. It is an associative, but clearly non-commutative product
on the class of suitably composable $SLH$ models.

\subsection{Fermion Fields}
In the above, we have set out the theory for Bosonic field inputs, however, in many applications
to quantum transport it would be natural to also consider fermionic fields. We are in the fortunate 
situation that the quantum stochastic calculus has a Fermionic version where we may consider 
anti-commuting fields $b^{in}_k (t), b^{in \dag}_k (t)$. The theory turns out to be structurally
identical to the Bose theory provided the $S$ and $H$ operators are even parity (so commute with the
fields) and the $L$ operators are odd (and so anti-commute with the fields.

The theory of Fermion quantum stochastic calculus is presented in \cite{HP_Fermi}.

\section{Quantum Transport Networks}

The standard component in quantum transport models is a device which may have
several contact points (or leads) which accept quantum field signals. For
definiteness, let us label the leads as $1,2,\cdots ,m$ and let $n_{k}$
denote the multiplicity of the $k$th lead. Our aim is to describe these
devices as quantum Markov models using the SLH formalism, and to develop
network rules to describe interconnected quantum transport components.

\begin{figure}[htbp]
	\centering
		\includegraphics[width=0.20\textwidth]{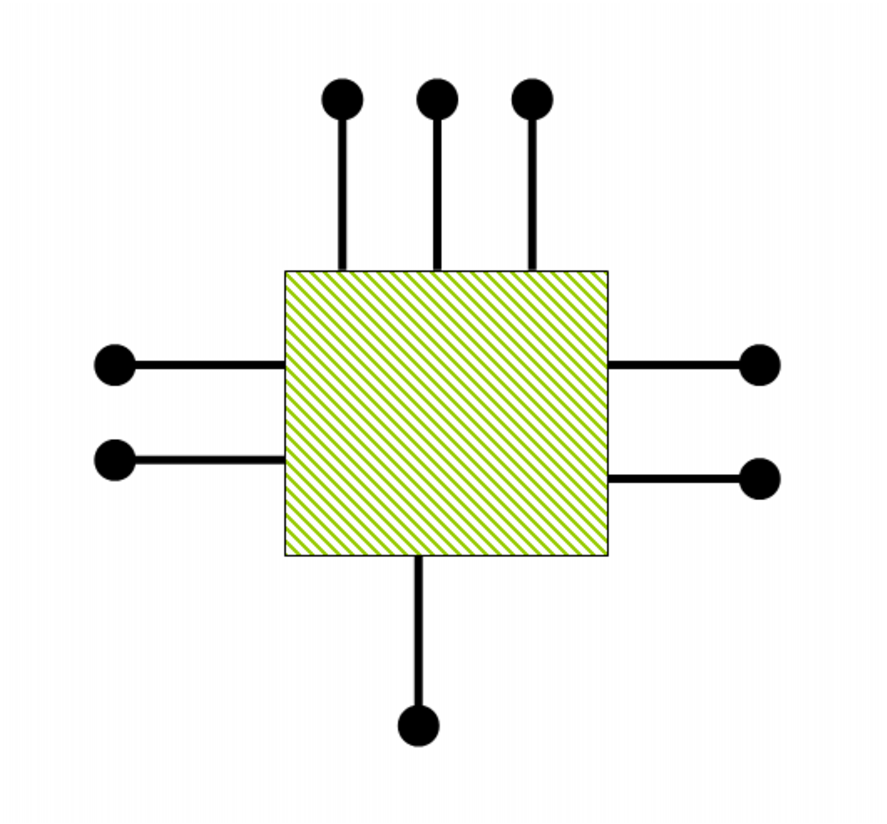}
	\caption{(color online) Single component with multiple lead contacts.}
	\label{fig:QTFN_device}
\end{figure}

\begin{figure}[htbp]
	\centering
		\includegraphics[width=0.450\textwidth]{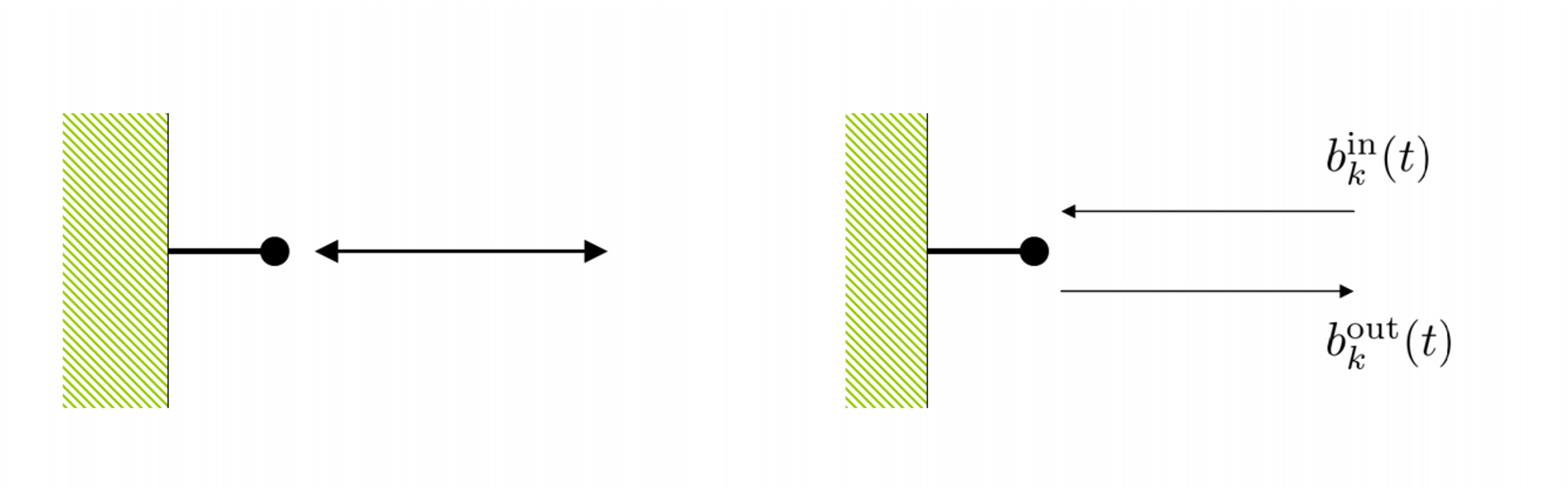}
	\caption{(color online) A bi-directional contact may be considered as an equivalent uni-directional input/output pair.}
	\label{fig:QTFN_contacts}
\end{figure}

The main difference between the quantum transport models and quantum
feedback networks is that in the former the fields are bi-directional while
in the latter they are uni-directional. This means that we may use the SLH
models to describe quantum transport components, but typically have to have
both an input and an output field to model each field terminating at a given
lead, see Fig. \ref{fig:QTFN_contacts}. 

As such the quantum transport models can be thought of as a special
form of SLH model, and their networks as a restricted class of quantum
feedback networks.

\begin{figure}[htbp]
	\centering
		\includegraphics[width=0.45\textwidth]{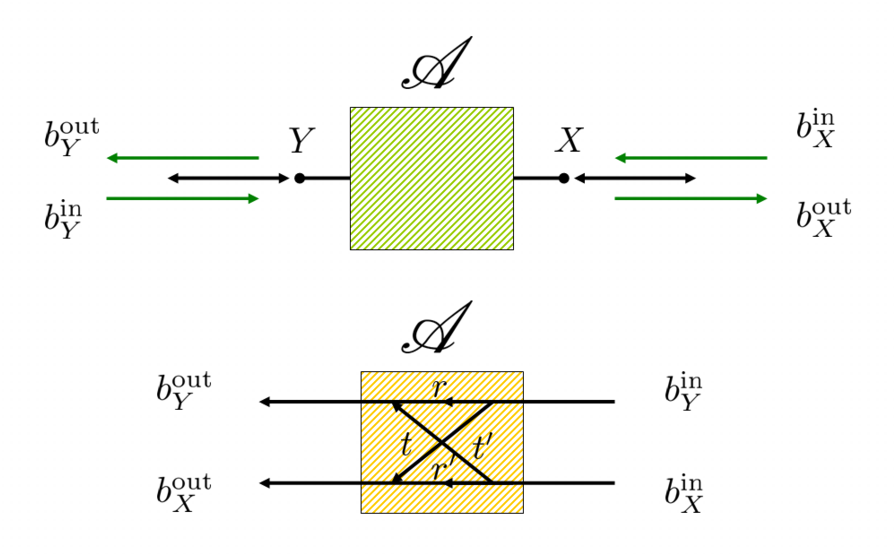}
	\caption{(color online) A two lead quantum transport device is naturally modelled as a 2 input - 2 output SLH component.}
	\label{fig:QTFN_equiv_SLH}
\end{figure}

A two-lead system is sketched in Fig. \ref{fig:QTFN_equiv_SLH} ( for simplicity we may assume that
each lead has multiplicity one, but this readily extends to multiple fields) and we formally identify this as a 2-input 2-output port SLH
system with 
\begin{equation*}
S=\left[ 
\begin{array}{cc}
S_{11} & S_{12} \\ 
S_{21} & S_{22}
\end{array}
\right] ,\;L=\left[ 
\begin{array}{c}
L_{1} \\ 
L_{2}
\end{array}
\right] ,\;H.
\end{equation*}

\begin{figure}[htbp]
	\centering
		\includegraphics[width=0.450\textwidth]{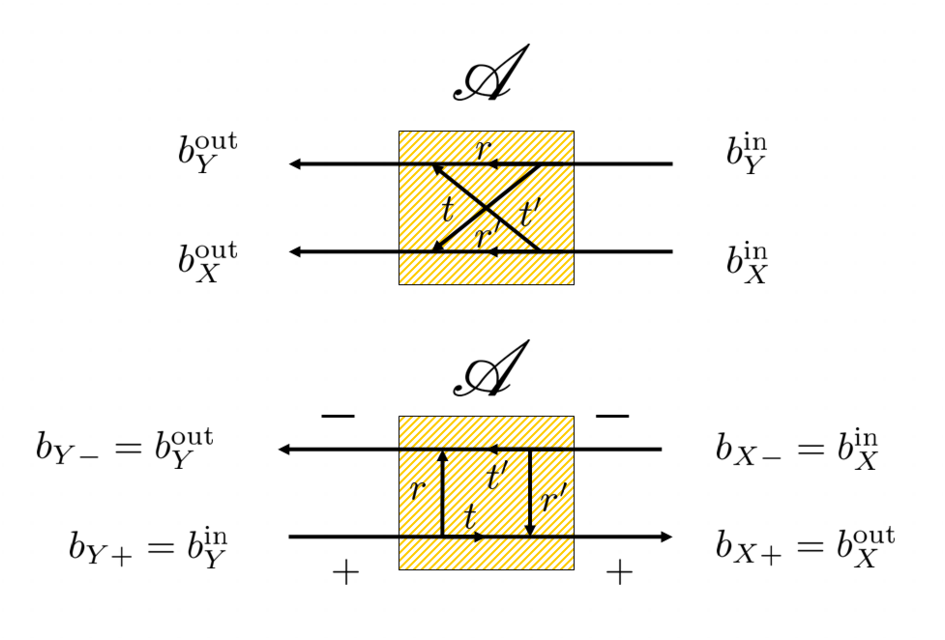}
	\caption{(color online) The usual input/output description is modified to have inputs and outputs corresponding to a given lead
	all appear grouped one one side.}
	\label{fig:QTFN_chain}
\end{figure}

The usual convention of displaying an SLH model, with all inputs on one side
and all outputs on the other, needs to be modified so that we end up with
input 1 and output 1 on one side and input 2 and output 2 on the other. The
transmission and reflection coefficients coefficients are listed in Fig. \ref{fig:QTFN_chain}
and we identify the matrix $S$ with the usual quantum transport scattering matrix as 
\begin{equation}
S =
\left[ 
\begin{array}{cc}
S_{YY} & S_{YX} \\ 
S_{XY} & S_{XX}
\end{array}
\right]
\equiv \left[ 
\begin{array}{cc}
r & t^{\prime } \\ 
t & r^{\prime }
\end{array}
\right] .
\end{equation}

\subsection{Quantum Transport Components in Series}

\begin{figure}[htbp]
\centering
\includegraphics[width=0.450\textwidth]{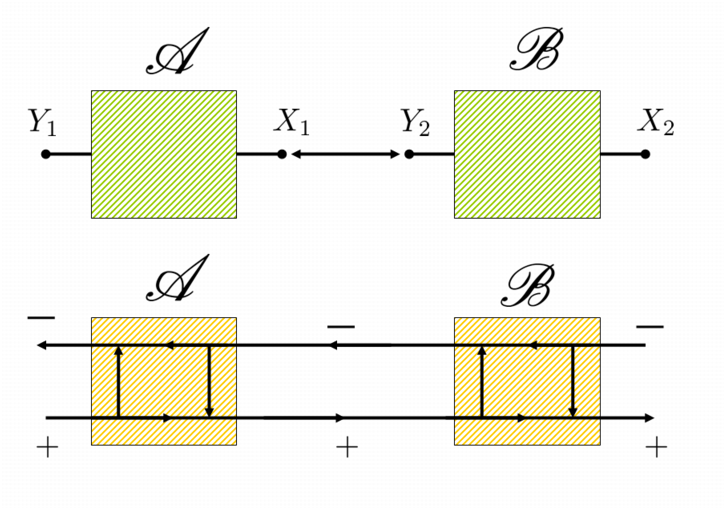}
\caption{(color online) A pair of cascaded quantum transport systems is
reinterpreted as a quantum feedback network.}
\label{fig:QTFN_chain_series}
\end{figure}

Our first step to build a network is to place two components, $\mathscr{A}$ and $\mathscr{B}$,
in series as shown in Fig. \ref{fig:QTFN_chain_series}. Here we connect the
quantum transmission line between two contact leads as indicated in the
upper part. In the SLH framework, we connect up the inputs and outputs as
shown in the lower part of Fig. \ref{fig:QTFN_chain_series}.

Let us take the SLH description of device $\mathscr{A}$ to be 
\begin{equation*}
G_{\mathscr{A}}\sim \left( S_{\mathscr{A}}=\left[ 
\begin{array}{cc}
S_{\mathscr{A}}^{-+} & S_{\mathscr{A}}^{--} \\ 
S_{\mathscr{A}}^{++} & S_{\mathscr{A}}^{+-}
\end{array}
\right] ,L_{\mathscr{A}}=\left[ 
\begin{array}{c}
L_{\mathscr{A}}^{-} \\ 
L_{\mathscr{A}}^{+}
\end{array}
\right] ,H_{\mathscr{A}}\right) ,
\end{equation*}

with a similar convention for $\mathscr{B}$. Here the indices $\pm $ indicate right
and left propagating noise fields.

\begin{figure}[h]
\centering
\includegraphics[width=0.450\textwidth]{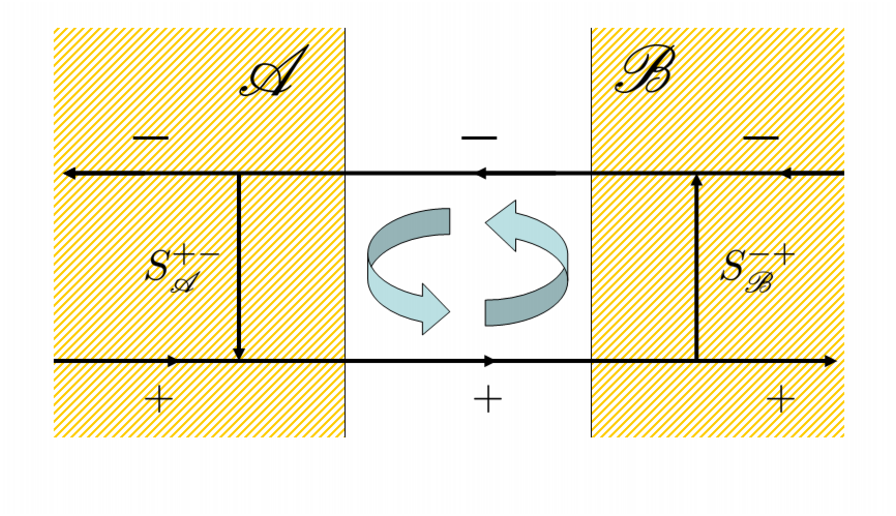}
\caption{(color online) The algebraic loop appearing in the cascaded quantum
transport set-up in Fig. \ref{fig:QTFN_chain_series}.}
\label{fig:QTFN_loop}
\end{figure}

The situation of two quantum transport systems in series differs
dramatically from the series product for uni-directional networks as now we
have the presence of an algebraic feedback loop, see Fig \ref{fig:QTFN_loop}.

In particular, we need the feedback reduction formula (\ref{eq:QFN_FR}) of 
\cite{GJ-QFN} to compute the resulting SLH. The construction is a \emph{%
Redheffer star product}, and is given by, see \cite{GJ-QFN} 
\begin{equation*}
G_{\mathscr{A}\bigstar \mathscr{B}}\sim \left( S_{\mathscr{A}\bigstar \mathscr{B}},L_{\mathscr{A}\bigstar \mathscr{B}},H_{\mathscr{A}\bigstar
\mathscr{B}}\right) 
\end{equation*}
where 
\begin{widetext}
\begin{eqnarray}
S_{\mathscr{A}\bigstar \mathscr{B}} &=&\left[ 
\begin{array}{cc}
S_{\mathscr{A}}^{-+}+S_{\mathscr{A}}^{--}W_{\mathscr{B}\mathscr{A}\mathscr{B}}^{-+}S_{\mathscr{B}}^{++} & S_{\mathscr{A}}^{--}Z_{\mathscr{B}\mathscr{A}}^{-}S_{\mathscr{B}}^{--}
\\ 
S_{\mathscr{B}}^{++}Z_{\mathscr{A}\mathscr{B}}^{+}S_{\mathscr{A}}^{++} & S_{\mathscr{B}}^{+-}+S_{\mathscr{B}}^{++}W_{\mathscr{A}\mathscr{B}\mathscr{A}}^{+-}S_{\mathscr{A}}^{--}
\end{array}
\right] ,\quad L_{\mathscr{A}\bigstar \mathscr{B}}=\left[ 
\begin{array}{c}
L_{\mathscr{A}}^{-}+S_{\mathscr{A}}^{--}Z_{\mathscr{B}\mathscr{A}}^{-}\left( L_{\mathscr{B}}^{-}+S_{\mathscr{B}}^{-+}L_{\mathscr{B}}^{+}\right) 
\nonumber
\\
L_{\mathscr{B}}^{+}+S_{\mathscr{B}}^{++}Z_{\mathscr{A}\mathscr{B}}^{+}\left( L_{\mathscr{A}}^{+}+S_{\mathscr{A}}^{+-}L_{\mathscr{B}}^{-}\right) 
\end{array}
\right] , \\
H_{\mathscr{A}\bigstar \mathscr{B}} &=&   H_{\mathscr{A}}+H_{\mathscr{B}}+\text{Im}\left\{ 
\begin{array}{c}
\left[ L_{\mathscr{A}}^{+\dag }+L_{\mathscr{B}}^{+\dag }S_{\mathscr{B}}^{++},L_{\mathscr{B}}^{-\dag }+L_{\mathscr{A}}^{-\dag
}S_{\mathscr{A}}^{--} 
 \right] \\  \,
\end{array}
 \left[ 
\begin{array}{cc}
Z_{\mathscr{A}\mathscr{B}}^{+} & W_{\mathscr{A}\mathscr{B}\mathscr{A}}^{+-} \\ 
W_{\mathscr{B}\mathscr{A}\mathscr{B}}^{-+} & Z_{\mathscr{B}\mathscr{A}}^{-}
\end{array}
\right] \left[ 
\begin{array}{c}
L_{\mathscr{A}}^{+} \\ 
L_{\mathscr{B}}^{-}
\end{array}
\right] \right\} ,
\label{eq:SLH_Redheffer}
\end{eqnarray}
\end{widetext}
with the following operators arising from the algebraic loop 
\begin{eqnarray*}
Z_{\mathscr{A}\mathscr{B}}^{+} &=&\left( 1-S_{\mathscr{A}}^{+-}S_{\mathscr{B}}^{-+}\right) ^{+1}, \\
Z_{\mathscr{B}\mathscr{A}}^{-} &=&\left( 1-S_{\mathscr{B}}^{-+}S_{\mathscr{A}}^{+-}\right) ^{+1}, \\
W_{\mathscr{A}\mathscr{B}\mathscr{A}}^{+-} &=&S_{\mathscr{A}}^{+-}Z_{\mathscr{B}\mathscr{A}}^{-}=Z_{\mathscr{A}\mathscr{B}}^{+}S_{\mathscr{A}}^{+-}, \\
W_{\mathscr{B}\mathscr{A}\mathscr{B}}^{-+} &=&S_{\mathscr{B}}^{-+}Z_{\mathscr{A}\mathscr{B}}^{+}=Z_{\mathscr{B}\mathscr{A}}^{-}S_{\mathscr{B}}^{-+}.
\end{eqnarray*}

\section{Quantum Linear Passive Markov Models}

It is convenient to assemble the inputs into a following column
vectors of length $n$
\begin{equation*}
\mathbf{b}^{\text{in}}\left( t\right) =\left[
\begin{array}{c}
b_{1}\left( t\right) \\
\vdots \\
b_{n}\left( t\right)
\end{array}
\right] .
\end{equation*}

The input-output relations may then be written more compactly as
$\mathbf{b}^{\text{out}}\left( t\right) = j_t (S) \,
\mathbf{b}^{\text{in}}\left( t\right) + j_t (L) $.

We now specialise to a linear model of a quantum mechanical system consisting of a family of
harmonic oscillators $\left\{ a_{j}:j=1,\cdots ,m\right\} $ with
canonical commutation relations $\left[ a_{j},a_{k}\right]
=0=\left[ a_{j}^{\dag },a_{k}^{\dag }\right] $ and $\left[
a_{j},a_{k}^{\dag }\right] =\delta _{jk} $. We collect into column
vectors:
\begin{equation}
\mathbf{a}=\left[
\begin{array}{c}
a_{1} \\
\vdots \\
a_{m}
\end{array}
\right] .
\end{equation}

Our interest is in the general linear open dynamical system and
this corresponds to the following situation:

\begin{itemize}
\item[1)]  The $S_{jk}$ are scalars.

\item[2)]  The $L_{j}^{\prime }s$ are linear, i.e., there exist constants $%
c_{jk}$ such that $L_{j}\equiv \sum_{k}c_{jk}a_{k}$.

\item[3)]  $H$ is quadratic, i.e., there exist constants $\omega
_{jk}$ such that $H=\sum_{jk}a_{j}^{\dag }\omega _{jk}a_{k}$.
\end{itemize}

The complex damping is $\frac{1}{2}L^{\dag }L+iH=-\mathbf{a}^{\dag }A\mathbf{%
a}$ where 
\begin{equation}
A=-\frac{1}{2}C^{\dag }C-i\Omega 
\end{equation}
with $C=\left(
c_{jk}\right) $ and $\Omega =\left( \omega _{jk}\right) $. Note
that $\Omega =\Omega ^{\dag
} $ because $H$ is self-adjoint, hence the real part of $A$ is $-\frac{1}{2}%
C^{\dag }C\leq 0$.

The Heisenberg-Langevin equations for $\mathbf{a}\left( t\right)
=j_t ( \mathbf{a} ) $ and the 
input-output relations then become
\begin{eqnarray*}
\mathbf{\dot{a}}\left( t\right) &=&A\mathbf{a}\left( t\right) -C^{\dag }S%
\mathbf{b}(t), \\
\mathbf{b}^{\text{out}}\left( t\right) &=&S\mathbf{b}\left( t\right) +C%
\mathbf{a}\left( t\right) .
\end{eqnarray*}
These linear equations are amenable to Laplace transform
techniques \cite {YK1},\cite{YK2}. We define for $\text{Re}s>0$
\begin{equation*}
{X}[s] =\int_{0}^{\infty }e^{-st}X\left( t\right)
dt,
\end{equation*}
where $X$ is now any of our stochastic processes. Note that ${%
\mathbf{\dot{a}}}[s] =s\mathbf{{a}}[s] -\mathbf{a%
}$. We find that
\begin{eqnarray*}
\mathbf{{a}}[s] &=&-\left( sI_{m}-A\right) ^{-1}C^{\dag }S%
\mathbf{{b}}^{\text{in}}[s] +\left( sI_{m}-A\right) ^{-1}%
\mathbf{a}, \\
\mathbf{{b}}^{\text{out}}[s] &=&S\mathbf{{b}}^{\text{in}%
}[s] +C\mathbf{{a}}[s] .
\end{eqnarray*}

The operator $\mathbf{{a}}[s] $ may be eliminated to give
\begin{equation}
\mathbf{{b}}^{\text{out}}[s] =\Xi (s) \,  \mathbf{{b}}^{\text{in}}[s] +\xi (s) \,
\mathbf{a}
\end{equation}
where the \textit{transfer matrix function} is
\begin{equation}
\Xi (s) \triangleq S-C\left( sI_{m}-A\right) ^{-1}C^{\dag }S
\label{eq:transfer}
\end{equation}
and $\xi [s] =C\left( sI_{m}-A\right) ^{-1}$.

If we average over the vacuum state of the environment, then we
would find that ${\frac{d }{dt}} \langle \mathbf{a} (t)
\rangle_{\text{vac} } =A \langle \mathbf{a} (t)
\rangle_{\text{vac} } $. The system is said to be internally
stable if $\langle \mathbf{a} (t) \rangle_{\text{vac} }
\rightarrow 0$ as $t\rightarrow \infty$. This occurs if and only
if $A$ is \textit{Hurwitz}, that is, all its eigenvalues have
negative real part.

As an example, consider a single mode cavity coupling to the input
field via $L=\sqrt{\gamma }a,$ and with Hamiltonian $H=\omega
a^{\dag }a$. This implies $A=-(\frac{\gamma }{2}+i\omega )$ and
$C=\sqrt{\gamma }$. If the output picks up an additional phase
$S=e^{i\phi }$, the corresponding transfer function is then
computed to be
\begin{equation}
\Xi _{cavity}\left( s\right) =e^{i\phi }\,\frac{s+i\omega -\frac{\gamma }{2}%
}{s+i\omega +\frac{\gamma }{2}}.
\end{equation}

For a single mode $a$ with two inputs $b_1^{\text{in}}$ and $b_2^{\text{in}}$, the choice 
\begin{equation*}
S=\left[ 
\begin{array}{cc}
1 & 0 \\ 
0 & 1
\end{array}
\right] ,C=\left[ 
\begin{array}{c}
\sqrt{\gamma _1} \\ 
\sqrt{\gamma _2}
\end{array}
\right] ,\Omega =\omega _{0}
\end{equation*}
describes the damped harmonic oscillator with unperturbed Hamiltonian $%
H=\omega _{0}a^{\dag }a$ and coupling operators $L_1=\sqrt{%
\gamma _1}a$ and $L_2=\sqrt{\gamma _2}a$
to the respective inputs. The transfer function is then 
\begin{eqnarray}
&&\Xi \left( s\right) =\frac{1}{s+\frac{1}{2}\left( \gamma _1+\gamma _2\right) +i\omega _{0}} \nonumber \\
&&\times \left[ 
\begin{array}{cc}
s-\frac{1}{2}\gamma _1+\frac{1}{2}\gamma _2+i\omega
_{0} & \sqrt{\gamma _1\gamma _2} \\ 
\sqrt{\gamma _1\gamma _2} & s+\frac{1}{2}\gamma _1 -\frac{1}{2}\gamma _2+i\omega _{0}
\end{array}
\right] . \quad
\label{eq:double}
\end{eqnarray}
\bigskip

The models are therefore determined completely by the matrices $\left( S,C,\Omega \right) $ with $S\in \mathbb{C}^{n\times n},C\in \mathbb{C}%
^{n\times m}$ and $\Omega \in \mathbb{C}^{m\times m}$ which of course give the SLH coefficients. We shall
use the convention $\left[
\begin{tabular}{l|l}
$A$ & $B$ \\ \hline
$C$ & $D$%
\end{tabular}
\right] (s) =D+C\left( s-A\right) ^{-1}B$ for matrices
$A\in \mathbb{C}^{m\times m},B\in \mathbb{C}^{m\times n},C\in
\mathbb{C}^{n\times m}$ and $D\in \mathbb{C}^{n\times n}$, and
write the transfer matrix function as
\begin{equation}
\Xi (s) =\left[
\begin{tabular}{r|r}
$A$ & $-C^{\dag }S$ \\ \hline
$C$ & $S$%
\end{tabular}
\right] (s)  ,  \label{TF}
\end{equation}
where $A=-\frac{1}{2}C^{\dag }C-i\Omega $. We note the
decomposition
\begin{equation*}
\Xi =\left[ I_{n}-C\left( sI_{m}-A\right) ^{-1}C^{\dag }\right]
S\equiv \left[
\begin{tabular}{r|r}
$A$ & $-C^{\dag }$ \\ \hline
$C$ & $I_{n}$%
\end{tabular}
\right] S.
\end{equation*}

\noindent
\textbf{Lemma 1} (All-pass representation of $\Xi $.) We may write the
transfer function $\Xi $ for a passive linear quantum system as
\begin{equation}
\Xi \left( s\right) =\frac{1-\frac{1}{2}\Sigma \left( s\right) }{1+\frac{1}{2%
}\Sigma \left( s\right) }S  \label{eq:allpass}
\end{equation}
where 
\begin{equation}
\Sigma \left( s\right) =C\frac{1}{s+i\Omega }C^{\dag }.
\end{equation}

\noindent
\textit{proof:} From the Woodbury matrix identity we find
\begin{multline*}
\frac{1}{s+\frac{1}{2}C^{\dag }C+i\Omega }=\\
\frac{1}{s+i\Omega } 
-\frac{1}{2}\frac{1}{s+i\Omega }C^\dag \frac{1}{1+\frac{1}{2}C\dfrac{1}{s+i\Omega 
}C^{\dag }}C \frac{1}{s+i\Omega },
\end{multline*}
which we substitute into (\ref{eq:transfer}) to get the result.
$\square$

\bigskip

\noindent
\textbf{Theorem 1} The transfer function of a passive system is inner, that
is $\Xi \left( i\omega \right) $ is unitary for all real $\omega$ not an eigenvalue of $\Omega $.

\bigskip

\noindent
\textit{proof:} For $\omega $ not an eigenvalue of $\Omega $, $\Sigma \left(
i\omega \right) =-iC\frac{1}{\omega +\Omega }C^{\dag }$ is well-defined and we have 
$\Sigma \left( i\omega \right) ^{\dag }=-\Sigma \left( i\omega \right) $, so that
unitarity of $\Xi \left( i\omega \right) $ follows from (\ref{eq:allpass}).
$\square$

\bigskip

Transfer functions that are inner are otherwise referred to as all-pass transfer functions
as classically this means that harmonic signals of arbitrary frequency pass through without
attenuation. In the current context it relates the fact that the output processes are is again canonical
field processes.

\subsection{The Chain Scattering Representation}
We now consider a linear transformation 
\begin{equation}
\left[ 
\begin{array}{c}
z_{1} \\ 
z_{2}
\end{array}
\right] =K\left[ 
\begin{array}{c}
u_{1} \\ 
u_{2}
\end{array}
\right] \equiv \left[ 
\begin{array}{cc}
K_{12} & K_{12} \\ 
K_{21} & K_{22}
\end{array}
\right] \left[ 
\begin{array}{c}
u_{1} \\ 
u_{2}
\end{array}
\right]   \label{eq:io}
\end{equation}
where $u_{1},u_{2},z_{1},z_{2}$ are all column vectors of equal length. Our
aim is to rewrite this in the form 
\begin{equation}
\left[ 
\begin{array}{c}
z_{1} \\ 
u_{1}
\end{array}
\right] =\mathtt{CHAIN}\left( K\right) \left[ 
\begin{array}{c}
u_{2} \\ 
z_{2}
\end{array}
\right]   \label{eq:wave}
\end{equation}
which is possible if $K_{21}$ is invertible, in which case we have 
\begin{equation}
\mathtt{CHAIN}\left( K\right) \triangleq \left[ 
\begin{array}{cc}
K_{12}-K_{11}K_{21}^{-1}K_{22} \; \; & \; \; \; K_{12}K_{21}^{-1} \\ 
-K_{21}^{-1}K_{22} & K_{21}^{-1}
\end{array}
\right] ,
\end{equation}
with inverse transformation
\[
\mathtt{CHAIN}^{-1} \left( M\right) \triangleq \left[ 
\begin{array}{cc}
M_{12} M_{22}^{-1} ; \;   & \; \; \; M_{11}- M_{12} M_{22}^{-1} M_{21} \\ 
M_{22}^{-1}       & M_{22}^{-1} M_{21}
\end{array}
\right]
.
\]

The linear system (\ref{eq:io}) is an input-output representation, while the
new system (\ref{eq:wave}) is called the chain-scattering representation. In
the former we have a state-based model where the system is driven by the inputs $u_{1},u_{2}$ 
and produces the outputs $z_{1},z_{2}$, while in the latter the system is a wave 
scatterer from the
wave $u_{2},z_{2}$ at port 2 to the wave $z_{1},u_{1}$ at port 1.

In the case where $K$ is unitary, we have that 
\begin{equation*}
\left| u_{1}\right| ^{2}+\left| u_{2}\right| ^{2}=\left| z_{1}\right|
^{2}+\left| z_{2}\right| ^{2}
\end{equation*}
and rearranging gives 
\begin{equation*}
\left| u_{1}\right| ^{2}-\left| z_{1}\right| ^{2}=\left| z_{2}\right|
^{2}-\left| u_{2}\right| ^{2}.
\end{equation*}
This suggests that if $K$ implements a unitary transformation for field
inputs, then $\mathtt{CHAIN}\left( K\right) $ is implements a Bogoliubov
transformation. We shall establish this fact next.

\subsubsection{Invariance Symmetries of Chain Pairs}

\noindent \textbf{Definition} The $\flat $-conjugation is defined on
matrices of dimension $2n$ by 
\begin{equation*}
X^{\flat }=J_{n}X^{\dag }J_{n}
\end{equation*}
where 
\begin{equation*}
J_{n}=\left[ 
\begin{array}{cc}
I_{n} & 0 \\ 
0 & -I_{n}
\end{array}
\right] .
\end{equation*}
We say that $X$ is a $\flat $-isometry, $\flat $-coisometry, if we have $%
X^{\flat }X=I_{2n}$, $XX^{\flat }=I_{2n}$, respectively. If $X$ is both a $%
\flat $-isometry and a $\flat $-coisometry, then we say that it is a $\flat $%
-unitary.

We now state the main structural properties of the chain scattering
transformation.

\bigskip 

\noindent \textbf{Theorem 2} A matrix $K$ is an isometry,
coisometry, unitary if and only if $M=\mathtt{CHAIN}\left( K\right) $ is a $\flat $-isometry. $\flat $-coisometry, $\flat $-unitary, respectively

\bigskip

The proof is quite cumbersome and we relegate to the Appendix.

\subsection{Wave Scattering in Quantum Transport}
It is convenient to relabel the input and output fields (and their Laplace transforms)
appearing in the upper
picture in Fig. \ref{fig:QTFN_chain} as 
\begin{eqnarray*}
b_{X+} &=& b_X^{\text{out}}   \\
b_{X-} &=& b_X^{\text{in}}   \\
b_{Y+} &=& b_Y^{\text{in}}   \\
b_{Y-} &=& b_Y^{\text{out}}   
\end{eqnarray*}
where the subscripts $+/-$ now indicate right and left propagating fields respectively.
The relation between these fields is then
\begin{equation*}
\left[ 
\begin{array}{c}
b_{Y-} \\ 
b_{X+}
\end{array}
\right] =\Xi \, \left[ 
\begin{array}{c}
b_{Y+} \\ 
b_{X-}
\end{array}
\right] \equiv \left[ 
\begin{array}{cc}
\Xi _{YY}^{-+} & \Xi _{YX}^{--} \\ 
\Xi _{XY}^{++} & \Xi _{XX}^{+-}
\end{array}
\right] \,
\left[ 
\begin{array}{c}
b_{Y+} \\ 
b_{X-}
\end{array}
\right]
  .
\end{equation*}
where we break down the transfer matrix into block form. We now perform the
swap described in the lower picture in Fig. \ref{fig:QTFN_chain}.

The equations may be rearranged as
\begin{equation}
\left[ 
\begin{array}{c}
b_{Y-}\\ 
b_{Y+}
\end{array}
\right] 
 =\left[ 
\begin{array}{cc}
\Gamma _{YX}^{--} & \Gamma _{YX}^{-+} \\ 
\Gamma _{YX}^{+-} & \Gamma _{YX}^{++}
\end{array}
\right] \,
\left[ 
\begin{array}{c}
b_{X-}\\ 
b_{X+}
\end{array}
\right] .  \label{eq:chain_scat_rep}
\end{equation}
or equivalently 
\begin{equation*}
\overleftrightarrow{b_Y} 
=\Gamma _{YX} \,  \overleftrightarrow{b_X} ,
\end{equation*}
where we introduce the following shorthand notation
for the inputs and outputs at a contact lead $X$
\begin{equation}
\overleftrightarrow{b_X} \triangleq
\left[ 
\begin{array}{c}
b_{X-}\\ 
b_{X+}
\end{array}
\right] .
\end{equation}

It immediately follows that
\begin{eqnarray}
\Gamma _{YX} (s) \equiv \mathtt{CHAIN}\left( \Xi (s)\right) 
\label{eq:Gamma=ChainXi}
\end{eqnarray}
that is 
\begin{eqnarray*}
\Gamma_{XY} =
\left[ 
\begin{array}{cc}
\Xi _{YX}^{--}-\Xi _{YY}^{-+} (\Xi _{XY}^{++}) ^{-1}\Xi _{XX}^{+-} \; \;  & \; \;\; \Xi _{YY}^{-+} (\Xi  _{XY}^{++}) ^{-1}\\ 
-(\Xi _{XY}^{++})^{-1}\Xi _{XX}^{+-} & (\Xi _{XY}^{++})^{-1}
\end{array}
\right] 
\end{eqnarray*}

Inversely, we have 
\begin{equation*}
\Xi =\left[ 
\begin{array}{cc}
\Gamma _{YX}^{-+}\left( \Gamma _{YX}^{++}\right) ^{-1} \; \; \;  & \;\ \; \; \; \Gamma
_{YX}^{--}-\Gamma _{YX}^{-+}\left( \Gamma _{YX}^{++}\right) ^{-1}\Gamma
_{YX}^{+-} \\ 
 \left( \Gamma
_{YX}^{++}\right) ^{-1} & -\left( \Gamma _{YX}^{++}\right) ^{-1}\Gamma _{YX}^{+-}
\end{array}
\right] 
\end{equation*}

\bigskip 

\noindent \textbf{Theorem 3}
The function $\Gamma_{YX} (s) =\mathtt{CHAIN}(\Xi (s))$  is a $\flat$-unitary 
for $s$ on the imaginary axis, except for eigenvalues of $-i\Omega$.

\bigskip

This is an immediate corollary to Theorems 1 and 2.

\bigskip

Note that the transfer function $\Gamma _{YX}$ connects the inputs and
outputs at contact lead $X$ to those at $Y$. As before, $\Gamma
_{YX}^{--}$ is a Schur complement of the transfer function $\Xi $ in block
matrix form. We shall always suppose that the chain-scattering
representation is valid, that is $\Xi _{XY}^{++}$ is invertible so that $\Gamma
_{YX}$ is well-defined.

\subsubsection{Chain-Scattering for Quantum Transport Devices in Series}
\begin{figure}[htbp]
	\centering
		\includegraphics[width=0.450\textwidth]{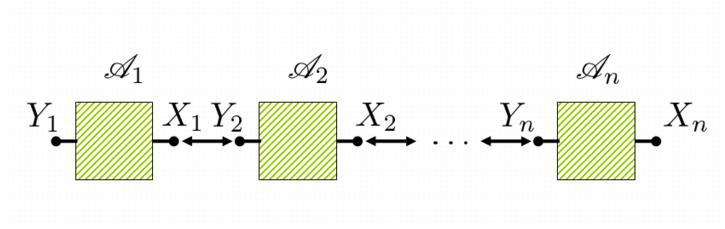}
	\caption{(color online) Several quantum transport components in a chain series.}
	\label{fig:QTFN_long_chain}
\end{figure}

Let us return to the devices in series shown in the upper diagram in
Fig. \ref{fig:QTFN_chain_series}. 
We have 

\begin{equation*}
\overleftrightarrow{b_{Y_1}} 
=\Gamma _{Y_1 X_1} \, \overleftrightarrow{b_{X_1}},
\quad \overleftrightarrow{b_{Y_2}}
=\Gamma _{Y_2 X_2 } \, \overleftrightarrow{b_{Y_2} } ,
\end{equation*}
however the identification
$b_{Y_2, +} \equiv b_{X_1 ,+}$ and $b_{Y_2, -}  \equiv b_{X_1, -}$ (i.e., $\overleftrightarrow{b_{Y_2}} \equiv \overleftrightarrow{b_{X_1}}$)
now implies that
\begin{eqnarray}
\overleftrightarrow{b_{Y_1}} =
\Gamma_{Y_1 X_1} \Gamma_{Y_2 X_2} \,
\overleftrightarrow{b_{Y_2}} .
\label{eq:chain_series}
\end{eqnarray}
The general rule is easy to state at this stage. For the chain of
components shown in Fig. \ref{fig:QTFN_long_chain} we have
\begin{eqnarray}
\Gamma_{Y_1 X_n}= \Gamma_{Y_1 X_1} \Gamma_{Y_2 X_2} \cdots  \Gamma_{Y_n X_n}.
\end{eqnarray}

\subsubsection{Coprime Factorisations}

We say that transfer function $\Gamma _{YX}$ has a factorisation if we may
write it as
\begin{equation*}
\Gamma _{YX}=\Upsilon _{Y}^{-1}\Upsilon _{X}
\end{equation*}
and in this way we may write the lead-to-lead equations in a more symmetric
form as
\begin{equation*}
\Upsilon _{X}\overleftrightarrow{\,b_{X}}=\Upsilon _{Y}\overleftrightarrow{%
\,b_{Y}}.
\end{equation*}

So far we have not done anything particularly useful, however, we could ask
for more properties of the factorisation.

Let $H_{\infty }$ denote the set of Hardy functions, that is, the class of
complex-matrix valued functions $M\left( s\right) $ that are analytic in the
closed right hand complex plane (Re $s\geq 0$) with the property that the
limit values $M\left( i\omega +0^{+}\right) $ exist for almost all $\omega
\in \mathbb{R}$ and there is a finite upper bound on the largest singular
value of $M\left( s\right) $ over Re $s\geq 0$. 

A factorisation $\Gamma _{YX}=\Upsilon _{Y}^{-1}\Upsilon _{X}$ will be useful for 
control and design purposes if both $\Upsilon _{Y}$ and $ \Upsilon _{X}$ are 
rational functions in the Hardy class with the property that they have no
common zeroes on the closed right hand plane, including $s=\infty$.
The appropriate definition from control theory is given below, see for instance \cite{Paganini}.

\bigskip 

\noindent\textbf{Definition:} A pair of matrix valued functions $\Upsilon _{X}$ and $%
\Upsilon _{Y}$ are left coprime if there exists a pair of rational matrix
functions $Q,P$  in the Hardy class such that
\begin{equation*}
\Upsilon _{Y}P-Q\Upsilon _{X}=I.
\end{equation*}
A left coprime factorisation of rational proper function $\Gamma _{YX}$ is a
factorisation $\Gamma _{YX}=\Upsilon _{Y}^{-1}\Upsilon _{X}$ where $\Upsilon
_{X}$ and $\Upsilon _{Y}$ are left coprime, with $\Upsilon _{Y}^{-1}$ proper.

\bigskip 

An important property of the chain-scattering
representation is that
\begin{eqnarray}
\Gamma _{YX}
=
\left[ 
\begin{array}{cc}
I & - \Xi _{YY}^{-+}  \\ 
0 & \Xi _{XY}^{++}
\end{array}
\right] ^{-1}\left[ 
\begin{array}{cc}
\Xi _{YX}^{--} & 0 \\ 
-\Xi _{XX}^{+-} & I
\end{array}
\right] 
\end{eqnarray}
and if the original transfer function $\Xi$ is stable this corresponds to a left coprime factorisation of $\Gamma _{YX}$. See Kimura, section 4.1 
\cite{Kimura}. A right coprime factorisation is given by
\begin{eqnarray*}
\Gamma _{YX} =\left[ 
\begin{array}{cc}
\Xi _{YX}^{--} & \Xi _{YY}^{-+}  \\ 
0 & I
\end{array}
\right] \left[ 
\begin{array}{cc}
I & 0 \\ 
\Xi _{XX}^{+-} & \Xi _{XY}^{++}
\end{array}
\right] ^{-1} .
\end{eqnarray*}

\subsubsection{Worked Example}

We study a simple device corresponding to a single mode $a$
with a pair of contact leads $X$ and $Y$, both of which have one input and
one output field. Here we assume that the device scatters the input fields like a beam splitter.
but also is damped by these inputs, as well as undergoing its own harmonic frequency $\omega_0$.
In the Heisenberg-Langevin picture we consider dynamical
equations
\begin{eqnarray*}
\frac{d}{dt}a\left( t\right)  &=&-\left( \gamma +i\omega _{0}\right) a\left(
t\right) -\sqrt{\gamma }b_{X}^{\text{in}}(t)-\sqrt{\gamma }b_{Y}^{\text{in}%
}(t), \\
b_{X}^{\text{out}}(t) &=&\frac{1}{\sqrt{2}}b_{X}^{\text{in}}(t)-\frac{1}{%
\sqrt{2}}b_{Y}^{\text{in}}(t)+\sqrt{\gamma }a(t), \\
b_{X}^{\text{out}}(t) &=&\frac{1}{\sqrt{2}}b_{X}^{\text{in}}(t)+\frac{1}{%
\sqrt{2}}b_{Y}^{\text{in}}(t)+\sqrt{\gamma }a(t).
\end{eqnarray*}
which corresponds to the choice
\begin{equation*}
S=\frac{1}{\sqrt{2}}\left[ 
\begin{array}{cc}
1 & -1 \\ 
1 & 1
\end{array}
\right] ,\quad L=\left[ 
\begin{array}{c}
\sqrt{\gamma }a \\ 
\sqrt{\gamma }a
\end{array}
\right] ,\quad H=\omega _{0}a^{\dag }a,
\end{equation*}
that is, $C=\sqrt{\gamma }\left[ 
\begin{array}{c}
1 \\ 
1
\end{array}
\right] $ and $\Omega =\omega _{0}$. The transfer function is then
\begin{equation*}
\Xi \left[ s\right] =\frac{1}{\sqrt{2}}\left[ 
\begin{array}{cc}
\Theta \left( s\right)  & -1 \\ 
\Theta \left( s\right)  & 1
\end{array}
\right] 
\end{equation*}
with $\Theta \left( s\right) =\frac{s-\gamma +i\omega _{0}}{s+\gamma
+i\omega _{0}}$. Using the chain transformation (\ref{eq:Gamma=ChainXi}) we find that
\begin{equation*}
\Gamma _{YX}=\left[ 
\begin{array}{cc}
-\sqrt{2} & 1 \\ 
-1/\Theta \left( s\right)  & \sqrt{2}/\Theta \left( s\right) 
\end{array}
\right] \allowbreak 
\end{equation*}
which admits the coprime factorisation $\Gamma _{YX}=\Upsilon
_{Y}^{-1}\Upsilon _{X}$ with 
\begin{equation*}
\Upsilon _{Y}=\left[ 
\begin{array}{cc}
1 & -\frac{1}{\sqrt{2}}\Theta \left( s\right)  \\ 
0 & \frac{1}{\sqrt{2}}\Theta \left( s\right) 
\end{array}
\right] ,\quad \Upsilon _{X}=\left[ 
\begin{array}{cc}
-\frac{1}{\sqrt{2}} & 0 \\ 
-\frac{1}{\sqrt{2}} & 1
\end{array}
\right] .
\end{equation*}

\subsection{Stability and the Lossless Property}

We have seen from Theorem 1 that the transfer function $\Xi $ of a linear
passive quantum system is inner (unitary almost everywhere on the imaginary
axis). In addition, if the system is stable, that is, the matrix $A$
appearing in the state-based model is Hurwitz, then following control
theoretic terminology we say that the system $\Xi $ is lossless. For
lossless systems, we have that
\begin{equation*}
\Xi ^{\dag }\left( s\right) \Xi \left( s\right) \leq I
\end{equation*}
in the closed right hand complex plane.

Similarly, we say that a chain scattering transfer function $\Gamma $ is $%
\flat $-lossless if 
\begin{equation*}
\Gamma ^{\flat }\left( s\right) \Gamma \left( s\right) \leq I,
\end{equation*}
for all Re$s\geq 0$.

Generally speaking, connecting an assembly of stable components into a
network may result in marginal instability. For instance, some marginal
stability may arise, which in quantum devices correspond to decoherence free
subspace which may be of importance in designing quantum memory storage. It
is imperative to know when a given system is lossless. Fortunately, the two
notions of losslessness above coincide.

\bigskip 

\textbf{Theorem 4:} $\Xi $ is $\flat $-lossless if and only if it takes the
form $\Gamma =\mathtt{CHAIN}\left( \Xi \right) $ where $\Xi $ is lossless.

\bigskip 

This is proved as Lemma 4.4 of Kimura's book \cite{Kimura}.

\subsubsection{State Space Realisations}

If we have the triple $\left( S,C,\Omega \right) $ of the form
\begin{equation*}
S=\left[ 
\begin{array}{cc}
S_{YY} & S_{YX} \\ 
S_{XY} & S_{XX}
\end{array}
\right] ,C=\left[ 
\begin{array}{c}
C_{Y} \\ 
C_{X}
\end{array}
\right] ,
\end{equation*}
leading to the transfer function
\begin{equation*}
\Xi =\left[ 
\begin{tabular}{l|ll}
$A$ & $B_{Y}$ & $B_{X}$ \\ \hline
$C_{Y}$ & $S_{YY}$ & $S_{YX}$ \\ 
$C_{X}$ & $S_{XY}$ & $S_{XX}$%
\end{tabular}
\right] 
\end{equation*}
where $A=-\frac{1}{2}C_{X}^{\dag }C_{X}-\frac{1}{2}C_{Y}^{\dag
}C_{Y}-i\Omega $ and $B_{k}=-\sum_{j}C_{j}^{\dag }S_{jk}$. It follows that
\begin{eqnarray*}
&&\Gamma _{YX}\equiv \\
&& \left[ 
\begin{tabular}{l|ll}
$A-B_{Y}S_{XY}^{-1}C_{X}$ & $B_{X}-B_{Y}S_{XY}^{-1}S_{XX}$ & $%
B_{Y}S_{XY}^{-1}$ \\ \hline
$C_{Y}-S_{YY}S_{XY}^{-1}C_{X}$ & $S_{YX}-S_{YY}S_{XY}^{-1}S_{XX}$ & $%
S_{YY}S_{XY}^{-1}$ \\ 
$-S_{XY}^{-1}C_{X}$ & $-S_{XY}^{-1}S_{XX}$ & $S_{XY}^{-1}$%
\end{tabular}
\right] ,
\end{eqnarray*}
see Kimura \cite{Kimura} section 4.2.

\subsection{Feedback and Termination}

\begin{figure}[htbp]
	\centering
		\includegraphics[width=0.450\textwidth]{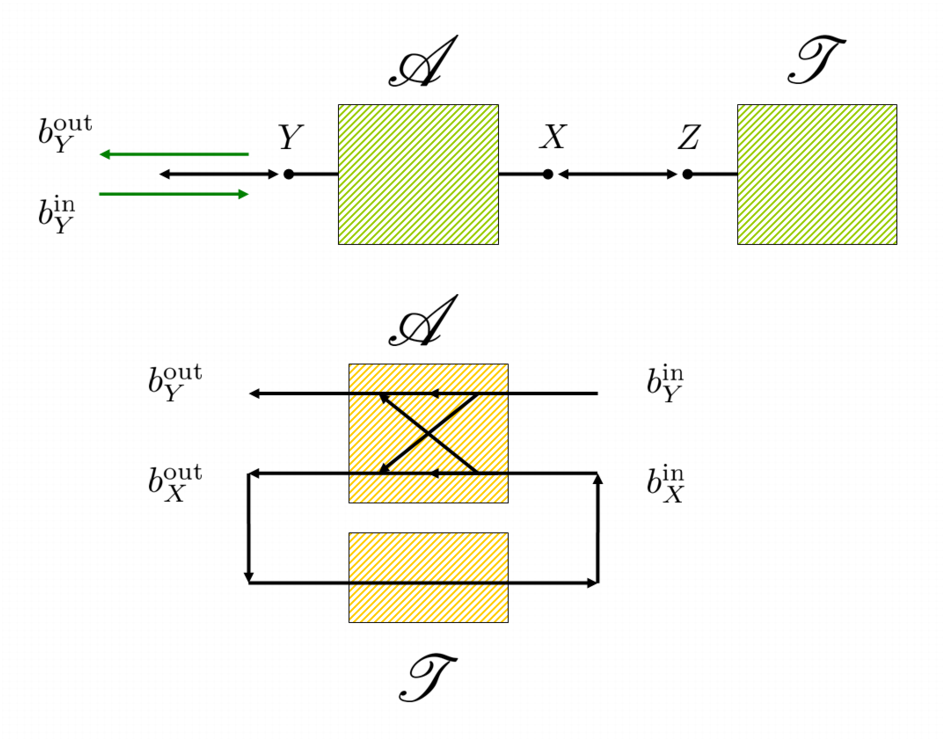}
	\caption{(color online) A standard procedure in circuit theory is to terminate a cascade of devices with a terminal load $\mathscr{T}$. In the input-output representation, this
	amounts to a feedback arrangement as shown.}
	\label{fig:QTFN_terminal1}
\end{figure}

We wish to consider now the effect of terminating a scattering sequence with
a terminal component $\Delta $, see Fig. \ref{fig:QTFN_terminal1}. The chain scattering picture
has a corresponding input-output representation where we see that the
component $\Delta $ is in fact in loop. With the identifications
\begin{equation}
b_{Y}^{\text{out}}=b_{Y-},\quad b_{Y}^{\text{in}}=b_{Y+}
\end{equation}
and from the relation $b_{X-}\left[ s\right] =\Delta \left( s\right) b_{X+}%
\left[ s\right] $, we may derive the input-output relation
\begin{equation*}
b_{Y-}\left[ s\right] =\Phi \left( s\right) \,b_{Y+}\left[ s\right] 
\end{equation*}
where $\Phi $ is a fractional linear transformation, see \cite{GGY},
\begin{equation}
\Phi \ =\Xi _{YY}^{-+}\ +\Xi _{YX}^{--}\ \Delta \ \left[ I-\Xi _{XX}^{+-}\
\Delta \right] ^{-1}\Xi _{XY}^{++}.
\end{equation}

We stress that the set-up in Fig. \ref{fig:QTFN_terminal1} is the special linear dynamical 
situation of the more general situation appearing in the feedback reduction rule Fig. \ref{fig:QFN_FR},
where we have the operator theoretic fractional linear transformation (\ref{eq:QFN_FR}). 

\begin{figure}[htbp]
	\centering
		\includegraphics[width=0.450\textwidth]{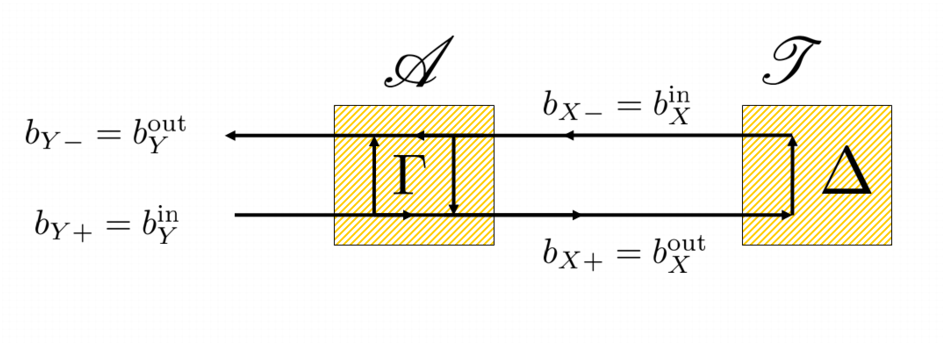}
	\caption{(color online) The chain scattering representation of the terminal load.}
	\label{fig:QTFN_terminal2}
\end{figure}

We may obtain a similar expression  in terms of the chain scattering
representation $\Gamma $, see Fig. \ref{fig:QTFN_terminal2}. Indeed we have
\begin{eqnarray*}
b_{Y-} &=&\left( \Gamma _{YX}^{--}\Delta +\Gamma _{YX}^{-+}\right) \,b_{X+},
\\
b_{Y+} &=&\left( \Gamma _{YX}^{+-}\Delta +\Gamma _{YX}^{++}\right) \,b_{X+},
\end{eqnarray*}
and so we deduce that
\begin{equation}
\Phi =\left( \Gamma _{YX}^{--}\Delta +\Gamma _{YX}^{-+}\right) \left( \Gamma
_{YX}^{+-}\Delta +\Gamma _{YX}^{++}\right) ^{-1}.
\end{equation}
This is equivalent to the homographic transformation from classical circuit
theory, and we write $\Phi \equiv \mathtt{HM}\left( \Gamma ,\Delta \right) $, 
following Kimura \cite{Kimura}.

\textbf{Theorem 5:} Let $\Xi $ be a quantum passive transfer functions
determined by $\left( \left[ 
\begin{array}{cc}
S_{YY} & S_{YX} \\ 
S_{XY} & S_{XX}
\end{array}
\right] ,\left[ 
\begin{array}{c}
C_{Y} \\ 
C_{X}
\end{array}
\right] ,\Omega \right) $ and let $\Delta =\left[ 
\begin{tabular}{l|l}
$A_{\Delta }$ & $B_{\Delta }$ \\ \hline
$C_{\Delta }$ & $D_{\Delta }$%
\end{tabular}
\right] $, where both representations are minimal \cite{minimal}. A minimal realisation of $
\Phi =\mathtt{HM}\left( \mathtt{CHAIN}\left( \Xi \right) ,\Delta \right) $
is given by
\begin{equation*}
\Phi =\left[ 
\begin{tabular}{l|l}
$A_{\Phi }$ & $B_{\Phi }$ \\ \hline
$C_{\Phi }$ & $D_{\Phi }$%
\end{tabular}
\right] 
\end{equation*}
where 
\begin{eqnarray*}
A_{\Phi } &=&\left[ 
\begin{array}{cc}
-\frac{1}{2}C^{\dag }C-i\Omega  & -C_{Y}^{\dag }C_{\Phi } \\ 
0 & A_{\Delta }
\end{array}
\right]  \\
&+& E_\Phi  \, \left( S_{XY}D_{\Delta }+S_{XX}\right) ^{-1}\left[ C_{X},\quad
S_{XY}C_{\Delta }\right]  \\
B_{\Phi } &=&E_\Phi \,  \left( S_{XY}D_{\Delta }+S_{XX}\right) ^{-1} \\
C_{\Phi } &=&\left[ C_{Y}-D_{\Phi }C_{X},\quad \left( S_{YY}-D_{\Phi
}S_{XY}\right) C_{\Delta }\right]  \\
D_{\Phi } &=&\left( S_{YY}D_{\Delta }+S_{YX}\right) \left( S_{XY}D_{\Delta
}+S_{XX}\right) ^{-1},
\end{eqnarray*}
where $E_\Phi$ is
\[
\left[
\begin{array}{c}
\left( C_{Y}^{\dag }S_{YY}+C_{X}^{\dag }S_{XY}\right) D_{\Delta }+\left(
C_{Y}^{\dag }S_{YX}+C_{X}^{\dag }S_{XX}\right)  \\ 
BS_{\Delta }
\end{array}
\right]
,
\]
provided that the network is well-posed (that is, the operator $S_{XY}S_{\Delta }+S_{XX}$
is invertible). Given $\Xi $ fixed, there will exist a $\Delta $ such that
the network is internally stable, that is, $A_{\Phi }$ is Hurwitz, if and
only if $\Xi $ is lossless.

\textit{proof:}
The state based representation derives form equations (4.84-4.87) of Kimura
with the explicit form of a quantum transfer function employed for $\Xi $.
The internal stability result follows from Theorem 4.15 of Kimura which
states that for $\flat $-unitary $\Gamma $, there will exist a $\Delta $
that make $\Phi =\mathtt{HM}\left( \Gamma ,\Delta \right) $ internally
stable if and only if $\Gamma $ is $\flat $-lossless, along with the
observation that $\Gamma =\mathtt{CHAIN}\left( \Xi \right) $ is
automatically $\flat $-lossless whenever $\Xi $ is lossless (see Theorem 4).
$\square$

Note that the theorem makes no claim that the stabilising $\Delta$ belongs to the class
of transfer function corresponding to a (active or passive) quantum transfer function,
only that it takes on a state-based model form. In favourable situations, this may be synthesis
as another quantum device, however it may entail using classical components.

\subsection{Time delays}

We now consider the network with delays in the transmission line. In the
case of linear circuits, this is easily modelled by the transfer function 
\begin{equation*}
\theta \left( s\right) =e^{-s\tau }
\end{equation*}
where $\tau $ is the time delay in the transmission line. A simple network of two connected
systems with delay is depicted in Fig. \ref{fig:QTFN_delay} below.

\begin{figure}[htbp]
	\centering
		\includegraphics[width=0.450\textwidth]{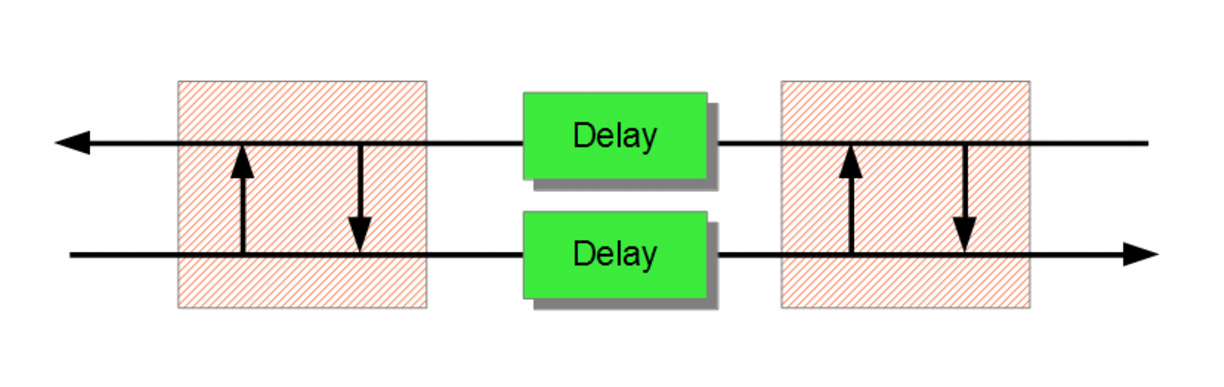}
	\caption{(color online) Cascaded system with delay.}
	\label{fig:QTFN_delay}
\end{figure}

In this case the
formula (\ref{eq:chain_series}) for the cascade of quantum transport models takes the form 
\begin{equation*}
\overleftrightarrow{b_{Y_{1}}}=\Gamma _{Y_{1}X_{1}}\,\Theta \,\Gamma
_{Y_{2}X_{2}}\,\overleftrightarrow{b_{X_{2}}}
\end{equation*}
where 
\begin{equation*}
\Theta =\left[ 
\begin{array}{cc}
\theta & 0 \\ 
0 & \theta ^{-1}
\end{array}
\right] .
\end{equation*}

\subsubsection{Trapped mode}
In the special case where we have only scattering, the transfer function
takes the form 
\begin{eqnarray*}
\Xi ^{-+} &=&r_{\mathscr{A}}+\theta ^{2}t_{\mathscr{A}}^{\prime }t_{%
\mathscr{A}}r_{\mathscr{B}}\frac{1}{1-\theta ^{2}r_{\mathscr{A}}r_{%
\mathscr{A}}^{\prime }}, \\
\Xi ^{--} &=&\theta t_{\mathscr{A}}^{\prime }t_{\mathscr{B}}^{\prime }\frac{1%
}{1-\theta ^{2}r_{\mathscr{A}}r_{\mathscr{A}}^{\prime }}, \\
\Xi ^{++} &=&\theta t_{\mathscr{A}}t_{\mathscr{B}}\frac{1}{1-\theta ^{2}r_{%
\mathscr{A}}r_{\mathscr{A}}^{\prime }}, \\
\Xi ^{+-} &=&r_{\mathscr{B}}^{\prime }+\theta t_{\mathscr{B}}t_{\mathscr{B}%
}^{\prime }r_{\mathscr{A}}^{\prime }\frac{1}{1-\theta ^{2}r_{\mathscr{A}}r_{%
\mathscr{A}}^{\prime }}.
\end{eqnarray*}

If we suppose that the transmittivity is weak of order with both $\left| t_{%
\mathscr{A}}\right| ^{2}$ and $\left| t_{\mathscr{B}}\right| ^{2}$ of order $%
\tau $, then we may obtain a well defined limit for small delay $\tau $. In
particular we set 
\begin{eqnarray*}
S_{\mathscr{A}} &=& \left[ 
\begin{array}{cc}
\sqrt{1-2\gamma _{\mathscr{A}}\tau } & -\sqrt{2\gamma _{\mathscr{A}}\tau }
\\ 
\sqrt{2\gamma _{\mathscr{A}}\tau } & \sqrt{1-2\gamma _{\mathscr{A}}\tau }
\end{array}
\right] ,\\
S_{\mathscr{B}} &=& \left[ 
\begin{array}{cc}
\sqrt{1-2\gamma _{\mathscr{B}}\tau } & -\sqrt{2\gamma _{\mathscr{B}}\tau }
\\ 
\sqrt{2\gamma _{\mathscr{B}}\tau } & \sqrt{1-2\gamma _{\mathscr{B}}\tau }
\end{array}
\right] ,
\end{eqnarray*}
where $\gamma _{\mathscr{A}}$ and $\gamma _{\mathscr{B}}$ are positive
constants. The limit $\tau \rightarrow 0$ leads to 
\begin{eqnarray*}
\lim_{\tau \rightarrow 0}\Xi \left( s,\tau \right) & =& \frac{1}{s+\frac{1}{2}%
\left( \gamma _{\mathscr{A}}+\gamma _{\mathscr{B}}\right) } \\
&& \times \left[ 
\begin{array}{cc}
s-\frac{1}{2}\gamma _{\mathscr{A}}+\frac{1}{2}\gamma _{\mathscr{B}} & \sqrt{%
\gamma _{\mathscr{A}}\gamma _{\mathscr{B}}} \\ 
\sqrt{\gamma _{\mathscr{A}}\gamma _{\mathscr{B}}} & s+\frac{1}{2}\gamma _{%
\mathscr{A}}-\frac{1}{2}\gamma _{\mathscr{B}}
\end{array}
\right] 
\end{eqnarray*}
which is the transfer function of a single mode with $\omega _{0}=0$ and two
input damping with rates $\gamma _{\mathscr{A}}$ and $\gamma _{\mathscr{B}}$, compare
with (\ref{eq:double}).

The limit corresponds to an effective trapped mode associated with the algebraic loop,
see for instance \cite{Bachor_Ralph} and \cite{GJN10}.

\section{Nonlinear Elements}

\begin{figure}[htbp]
	\centering
		\includegraphics[width=0.450\textwidth]{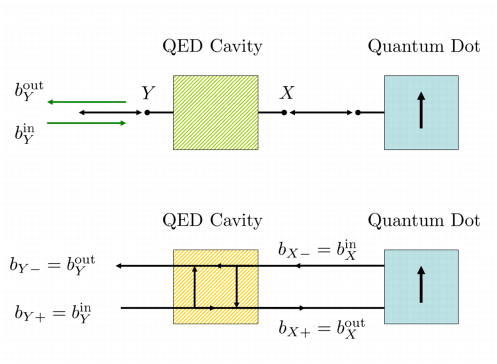}
	\caption{(color online) A cavity QED component is connected to a quantum dot (qubit system) so that
	the qubit is a nonlinear terminal load element. The chain scattering representation is sketched underneath.}
	\label{fig:QTFN_QD1}
\end{figure}

In our final example, we consider an nonlinear element in a quantum
transport network. Our example will consist of a quantum dot, modelled as
qubit system, acting as the terminal load of a network as shown in Fig. \ref{fig:QTFN_QD1}.

We take the SLH coefficients for the two-lead component device (a cavity QED
mode $a$) and the quantum dor to be respectively
\begin{eqnarray*}
G_{\text{QED}} \sim \left( \left[ 
\begin{array}{cc}
r & t^{\prime } \\ 
t & r^{\prime }
\end{array}
\right] ,\left[ 
\begin{array}{c}
\sqrt{\gamma _{+}}a \\ 
\sqrt{\gamma _{+}a}
\end{array}
\right] ,\omega _{0}a\right)  
\end{eqnarray*}
and 
\begin{eqnarray*}
G_{\text{QD}} \sim \left( e^{i\phi },\sqrt{\kappa }\sigma ,\omega ^{\prime
}\sigma _{z}\right) 
\end{eqnarray*}
where $\sigma $ is the lowering operator for the quantum dot qubit.

Following the network rules, we first form the parallel sum $G_{\text{QED}%
}\boxplus G_{\text{QD}}$ which has the model matrix
\begin{widetext}
\begin{equation*}
\mathsf{V}=\left[ 
\begin{array}{cccc}
K_{0} & -\left( \sqrt{\gamma _{+}}r+\sqrt{\gamma _{-}}t\right) a^{\dag } & 
-\left( \sqrt{\gamma _{+}}t^{\prime }+\sqrt{\gamma _{-}}r^{\prime }\right)
a^{\dag } & -e^{i\phi }\sqrt{\kappa }\sigma ^{\dag } \\ 
\sqrt{\gamma _{+}}a & r & t^{\prime } & 0 \\ 
\sqrt{\gamma _{-}}a & t & r^{\prime } & 0 \\ 
\sqrt{\kappa }\sigma  & 0 & 0 & e^{i\phi }
\end{array}
\right] 
\end{equation*}
\end{widetext}
with $K_{0}=-\frac{1}{2}\left( \gamma _{+}+\gamma _{-}\right) a^{\dag }a-%
\frac{1}{2}\kappa \sigma ^{\dag }\sigma -i\omega _{0}a^{\dag }a-i\omega
^{\prime }\sigma ^{\dag }\sigma $.

\begin{figure}[htbp]
	\centering
		\includegraphics[width=0.450\textwidth]{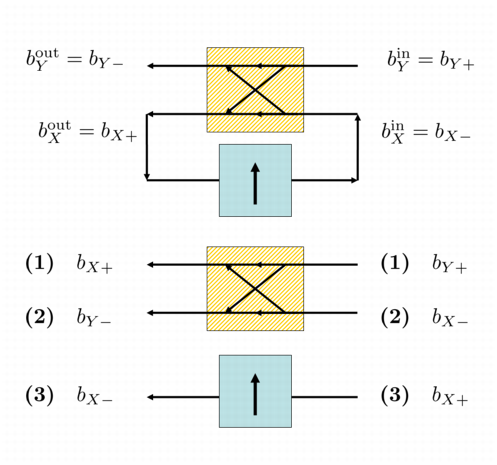}
	\caption{(color online) The equivalent quantum feedback network for the cavity-qubit system. Below it is the 
	open loop set-up before the feedback connections are made - note that the outputs and input are labelled so that output 1 goes to input 3, output 3 goes to input 2,
	while output 2 and input 1 are the external fields that remain after the feedback reduction.}
	\label{fig:QTFN_QD2}
\end{figure}

We now need to specify the connections we need to make: these are the pairs $%
\left( s,r\right) $ consiting of an output source $s$ and an input range $r$
label, and to wire up the open loop system Fig. \ref{fig:QTFN_QD2} these are $\left(
1,3\right) $ and $\left( 3,2\right) $. the corresponding adjacency matrix is
\begin{equation*}
\eta =\left[ 
\begin{array}{cc}
\eta _{12} & \eta _{13} \\ 
\eta _{32} & \eta _{33}
\end{array}
\right] =\left[ 
\begin{array}{cc}
0 & 1 \\ 
1 & 0
\end{array}
\right] 
\end{equation*}
where $\eta _{sr}=1$ if $\left( s,r\right) $ is a connection, and $=0$
otherwise. (The adjacency matrix ranges over the indices $s,r$ labelling the
outputs and inputs respectively that are to be connected to form the
feednback network.)

The feedback reduction formula (\ref{eq:QFN_FR}) now gives the closed loop network as
\begin{eqnarray*}
\mathscr{F} \left( \mathsf{V},\eta ^{-1}\right)  &=&\left[ 
\begin{array}{cc}
K_{0} & -\left( \sqrt{\gamma _{+}}r+\sqrt{\gamma _{-}}t\right) a^{\dag } \\ 
\sqrt{\kappa }\sigma  & 0
\end{array}
\right]  \\
&&+\left[ 
\begin{array}{cc}
-\left( \sqrt{\gamma _{+}}t^{\prime }+\sqrt{\gamma _{-}}r^{\prime }\right)
a^{\dag } & -e^{i\phi }\sqrt{\kappa }\sigma ^{\dag } \\ 
0 & e^{i\phi }
\end{array}
\right] \\
&&\left( \left[ 
\begin{array}{cc}
0 & 1 \\ 
1 & 0
\end{array}
\right] -\left[ 
\begin{array}{cc}
t^{\prime } & 0 \\ 
r^{\prime } & 0
\end{array}
\right] \right) ^{-1}\left[ 
\begin{array}{cc}
\sqrt{\gamma _{+}}a & r \\ 
\sqrt{\gamma _{-}}a & t
\end{array}
\right]  \\
&\equiv &\left[ 
\begin{array}{cc}
K & -L^{\dag }S \\ 
L & S
\end{array}
\right] 
\end{eqnarray*}
with $K=-\frac{1}{2}L^{\dag }L-iH$. After a little algebra we find the
equivalent SLH model to be
\begin{eqnarray}
S &=&e^{i\phi }\left( r+\frac{tt^{\prime }}{1-r^{\prime }}\right) ,  \notag
\\
L &=&e^{i\phi }\left( \sqrt{\gamma _{+}}+\frac{t^{\prime }}{1-r^{\prime }}%
\sqrt{\gamma _{-}}\right) a+\sqrt{\kappa }\sigma ,   \label{eq:L_QD} \\
K &=&-\left( \frac{\gamma _{+}+\gamma _{-}}{2}+\sqrt{\gamma _{-}}\frac{%
\left( \sqrt{\gamma _{+}}t^{\prime }+\sqrt{\gamma _{-}}r^{\prime }\right) }{%
1-r^{\prime }}+i\omega _{0}\right) a^{\dag }a \nonumber \\
&&-(\frac{1}{2}\kappa +i\omega ^{\prime })\sigma ^{\dag }\sigma -e^{i\phi }%
\sqrt{\kappa }\left( \sqrt{\gamma _{+}}+\frac{t^{\prime }}{1-r^{\prime }}%
\sqrt{\gamma _{-}}\right) \sigma ^{\dag }a  \nonumber \\ \label{eq:K_QD}
\end{eqnarray}
with the Hamiltonian $H$ determined as the solution to
\begin{eqnarray*}
-\frac{1}{2}L^{\dag }L-iH &=&K_{0}-\sqrt{\gamma _{-}}\left( \sqrt{\gamma _{+}%
}t^{\prime }+\sqrt{\gamma _{-}}r^{\prime }\right) \frac{1}{1-r^{\prime }}%
a^{\dag }a \\
&&-e^{i\phi }\sqrt{\kappa }\left( \sqrt{\gamma _{+}}+\frac{t^{\prime }}{%
1-r^{\prime }}\sqrt{\gamma _{-}}\right) \sigma ^{\dag }a.
\end{eqnarray*}

The input-output relation is then (with $b_{Y}^{\text{in}}=b_{Y+}$ and $%
b_{Y}^{\text{out}}=b_{Y-}$) 
\begin{eqnarray*}
b_{Y}^{\text{out}}(t) &=&e^{i\phi }\left( r+\frac{tt^{\prime }}{1-r^{\prime }%
}\right) b_{Y}^{\text{in}}(t) \\
&&+e^{i\phi }\left( \sqrt{\gamma _{+}}+\frac{t^{\prime }}{1-r^{\prime }}%
\sqrt{\gamma _{-}}\right) j_{t}(a)+\sqrt{\kappa }j_{t}(\sigma )
\end{eqnarray*}
where $j_{t}(a)$ and $j_{t}(\sigma )$ are the Heisenberg picture values of
the operators $a$ and $\sigma $. the master equation for joint density
states $\varrho $ of the QED cavity and qubit quantum dot are therefore
\begin{equation*}
\frac{d}{dt}\varrho =\frac{1}{2}L\varrho L^{\dag }-\varrho K^{\dag
}-K\varrho 
\end{equation*}
with $L$ and $K$ given by (\ref{eq:L_QD}) and (\ref{eq:K_QD}), respectively.

\section{Conclusion}
We have started the programme of developing a systematic network theory underlying
interconnections of quantum transport components in the direction that has proved successful so far for quantum
photonic networks. The existing quantum feedback network is shown to be capable of describing a large
class of nonlinear quantum transport components assembled into a network and is applicable to
modelling control design, especially as there is a growing interest in on-chip networks for
solid state quantum networks, and hybrid quantum transport-photonic circuits. We did not consider applications to control 
in this paper \textit{per se} but it is clear that many of the techniques currently used in quantum feedback
control for photinic networks are immediately applicable to this new domain.

\bigskip \textbf{Acknowledgement} 
The author has the pleasant duty to thank the Isaac Newton Institute for Mathematical Sciences, 
Cambridge, for support and hospitality during the programme \textit{Quantum Control Engineering} 
where work on this paper was completed. He is also acknowledges fruitful discussion with Clive Emary
on feedback control of quantum transport systems. 

\appendix

\section{proof of Theorem 2}

\label{sec:lemma1} The isometry condition $K^{\dag }K=I_{2n}$ implies the
identities 
\begin{equation*}
K_{1i}^{\dag }K_{1j}+K_{2i}^{\dag }K_{2j}=\delta _{ij}I_{n}.
\end{equation*}
Now 
\begin{equation*}
M^{\flat }=\left[ 
\begin{array}{cc}
M_{12}^{\dag } & -M_{22}^{\dag } \\ 
-M_{11}^{\dag } & M_{21}^{\dag }
\end{array}
\right] .
\end{equation*}
We establish the $\flat $-isometry property of $M$: (we collect together in
square brackets the various terms where we use the isometric property of $K$%
) 
\begin{eqnarray*}
&&\left[ M^{\flat }M\right] _{12} \\
&=&M_{12}^{\dag }M_{12}-M_{22}^{\dag }M_{22} \\
&=&\left( K_{12}^{\dag }-K_{22}^{\dag }K_{21}^{\dag -1}K_{11}^{\dag }\right)
\left( K_{12}-K_{11}K_{21}^{-1}K_{22}\right)  \\
&&-\left( K_{22}^{\dag }K_{21}^{\dag -1}\right) K_{21}^{-1}K_{22} \\
&=&K_{12}^{\dag }K_{12}-\left[ K_{12}^{\dag }K_{11}\right]
K_{21}^{-1}K_{22}-K_{22}^{\dag }K_{21}^{\dag -1}\left[ K_{11}^{\dag }K_{12}%
\right]  \\
&&+K_{22}^{\dag }K_{21}^{\dag -1}\left[ K_{11}^{\dag }K_{11}\right]
K_{21}^{-1}K_{22} \\
&&-K_{22}^{\dag }K_{21}^{\dag -1}K_{21}^{-1}K_{22} \\
&=&K_{12}^{\dag }K_{12}+\left( K_{22}^{\dag }K_{21}\right)
K_{21}^{-1}K_{22}+K_{22}^{\dag }K_{21}^{\dag -1}\left( K_{21}^{\dag
}K_{22}\right)  \\
&&+K_{22}^{\dag }K_{21}^{\dag -1}\left( I_{n_{2}}-K_{21}^{\dag
}K_{21}\right) K_{21}^{-1}K_{22} \\
&&-K_{22}^{\dag }K_{21}^{\dag -1}K_{21}^{-1}K_{22} \\
&=&K_{12}^{\dag }K_{12}+K_{22}^{\dag }K_{22}=I_{n_{1}};
\end{eqnarray*}

\begin{eqnarray*}
&&\left[ M^{\flat }M\right] _{11} \\
&=&M_{12}^{\dag }M_{11}-M_{22}^{\dag }M_{21} \\
&=&\left( K_{12}^{\dag }-K_{22}^{\dag }K_{21}^{\dag -1}K_{11}^{\dag }\right)
\left( K_{11}K_{21}^{-1}\right)  \\
&&-\left( -K_{22}^{\dag }K_{21}^{\dag -1}\right) \left( K_{21}^{-1}\right) 
\\
&=&\left[ K_{12}^{\dag }K_{11}\right] K_{21}^{-1}-K_{22}^{\dag }K_{21}^{\dag
-1}\left[ K_{11}^{\dag }K_{11}\right] K_{21}^{-1} \\
&&+K_{22}^{\dag }K_{21}^{\dag -1}K_{21}^{-1} \\
&=&\left( -K_{22}^{\dag }K_{21}\right) K_{21}^{-1}-K_{22}^{\dag
}K_{21}^{\dag -1}\left( I_{n_{2}}-K_{21}^{\dag }K_{21}\right) K_{21}^{-1} \\
&&+K_{22}^{\dag }K_{21}^{\dag -1}K_{21}^{-1}=0;
\end{eqnarray*}

\begin{eqnarray*}
&&\left[ M^{\flat }M\right] _{22} \\
&=&-M_{11}^{\dag }M_{12}+M_{21}^{\dag }M_{22} \\
&=&-\left( K_{21}^{\dag -1}K_{11}^{\dag }\right) \left(
K_{12}-K_{11}K_{21}^{-1}K_{22}\right)  \\
&&+K_{21}^{\dag -1}\left( -K_{21}^{-1}K_{22}\right)  \\
&=&-K_{21}^{\dag -1}\left[ K_{11}^{\dag }K_{12}\right] +K_{21}^{\dag -1}%
\left[ K_{11}^{\dag }K_{11}\right] K_{21}^{-1}K_{22} \\
&&-K_{21}^{\dag -1}K_{21}^{-1}K_{22} \\
&=&K_{21}^{\dag -1}K_{21}^{\dag }K_{22}+K_{21}^{\dag -1}\left[
I_{n_{2}}-K_{21}^{\dag }K_{21}\right] K_{21}^{-1}K_{22} \\
&&-K_{21}^{\dag -1}K_{21}^{-1}K_{22}=0;
\end{eqnarray*}
and 
\begin{eqnarray*}
\left[ M^{\flat }M\right] _{21} &=&-M_{11}^{\dag }M_{11}+M_{21}^{\dag }M_{21}
\\
&=&-\left( K_{21}^{\dag -1}K_{11}^{\dag }\right) \left(
K_{11}K_{21}^{-1}\right) +K_{21}^{\dag -1}K_{21}^{-1} \\
&=&-K_{21}^{\dag -1}\left[ K_{11}^{\dag }K_{11}\right] K_{21}^{-1}+K_{21}^{%
\dag -1}K_{21}^{-1} \\
&=&-K_{21}^{\dag -1}\left( I_{n_{2}}-K_{21}^{\dag }K_{21}\right)
K_{21}^{-1}+K_{21}^{\dag -1}K_{21}^{-1} \\
&=&I_{n_{2}}.
\end{eqnarray*}
Therefore $M^{\flat }M=I_{n}$ as required. The demonstration that the
coisometry of $K$ implies the $\flat $-coisometry of $M$ is similar.

We now establish the ``only if'' part of the theorem.

\label{sec:lemma2} We note that the $\flat $-isometry implies that 
\begin{eqnarray*}
M_{12}^{\dagger }M_{12}-M_{22}^{\dagger }M_{22} &=&I_{n_{1}}, \\
M_{21}^{\dagger }M_{21}-M_{11}^{\dagger }M_{11} &=&I_{n_{2}}, \\
M_{12}^{\dagger }M_{11}-M_{22}^{\dagger }M_{21} &=&0, \\
M_{21}^{\dagger }M_{22}-M_{11}^{\dagger }M_{12} &=&0
\end{eqnarray*}
and these imply respectively the following identities 
\begin{eqnarray}
I_{n_{1}} &=&K_{12}^{\dagger }K_{12}-K_{12}^{\dagger
}K_{11}K_{21}^{-1}K_{22}-K_{22}^{\dagger }K_{21}^{\dagger -1}K_{11}^{\dagger
}K_{12}  \notag  \label{B1} \\
&&+K_{22}^{\dagger }K_{21}^{\dagger -1}K_{11}^{\dagger
}K_{11}K_{21}^{-1}K_{22}-K_{22}^{\dagger }K_{21}^{\dagger
-1}K_{21}^{-1}K_{22},  \notag \\
&& \\
I_{n_{2}} &=&K_{21}^{\dagger -1}K_{21}^{-1}-K_{21}^{\dagger
-1}K_{11}^{\dagger }K_{11}K_{21}^{-1},  \label{B2} \\
0 &=&K_{12}^{\dagger }K_{11}K_{21}^{-1}-K_{22}^{\dagger }K_{21}^{\dagger
-1}K_{11}^{\dagger }K_{11}K_{21}^{-1}  \notag \\
&+&K_{22}^{\dagger }K_{21}^{\dagger -1}K_{21}^{-1}  \label{B3} \\
0 &=&-K_{21}^{\dagger -1}K_{21}^{-1}K_{22}-K_{21}^{\dagger -1}K_{11}K_{12} 
\notag \\
&+&K_{21}^{\dagger -1}K_{11}^{\dagger }K_{11}K_{21}^{-1}K_{22}  \label{B4}
\end{eqnarray}

To show the isometry property of $K$ we note that 
\begin{eqnarray}
&&K_{12}^{\dagger }K_{11}K_{21}^{-1}K_{22}  \notag \\
&\overset{\ref{B3}}{=}&\left[ K_{22}^{\dagger }K_{21}^{\dagger
-1}K_{11}^{\dagger }K_{11}K_{21}^{-1}-K_{22}^{\dagger }K_{21}^{\dagger
-1}K_{21}^{-1}\right] K_{22}  \notag \\
&=&K_{22}^{\dagger }\left[ K_{21}^{\dagger -1}K_{11}^{\dagger
}K_{11}K_{21}^{-1}-K_{21}^{\dagger -1}K_{21}^{-1}\right] K_{22}  \notag \\
&\overset{\ref{B2}}{=}&-K_{22}^{\dagger }K_{22}.  \label{B5}
\end{eqnarray}
Now we can compute the matrix elements: 
\begin{eqnarray*}
&&K_{12}^{\dagger }K_{12}+K_{22}^{\dagger }K_{22} \\
&\overset{\ref{B1}}{=}&I_{n_{1}}+K_{12}^{\dagger }K_{11}K_{21}^{-1}K_{22} \\
&&+K_{22}^{\dagger }K_{21}^{\dagger -1}K_{11}^{\dagger
}K_{12}+K_{22}^{\dagger }K_{22} \\
&&-K_{22}^{\dagger }K_{21}^{\dagger -1}K_{11}^{\dagger
}K_{11}K_{21}^{-1}M_{22}+K_{22}^{\dagger }K_{21}^{\dagger
-1}K_{21}^{-1}K_{22} \\
&=&I_{n_{1}}+K_{22}^{\dagger }\left[ -K_{21}^{\dagger -1}K_{11}^{\dagger
}K_{11}K_{21}^{-1}+K_{21}^{\dagger -1}K_{21}^{-1}\right] K_{22} \\
&&+K_{12}^{\dagger }K_{11}K_{21}^{-1}K_{22}+\left( K_{12}^{\dagger
}K_{11}K_{21}^{\dagger -1}K_{22}\right) ^{\dagger }+K_{22}^{\dagger }K_{22}
\\
&\overset{\ref{B2}}{=}&I_{n_{1}}+2K_{22}^{\dagger }K_{22}+K_{12}^{\dagger
}K_{11}K_{21}^{-1}K_{22} \\
&&+\left( K_{12}^{\dagger }K_{11}K_{21}^{-1}K_{22}\right) ^{\dagger } \\
&\overset{\ref{B5}}{=}&I_{n_{1}};
\end{eqnarray*}

\begin{eqnarray*}
&&K_{12}^{\dagger }K_{11}+K_{22}^{\dagger }K_{21} \\
&=&\left( K_{12}^{\dagger }K_{11}K_{21}^{-1}+K_{22}^{\dagger }\right) K_{21}
\\
&\overset{\ref{B3}}{=}&\left( K_{22}^{\dagger }K_{21}^{\dagger
-1}K_{11}^{\dagger }K_{11}K_{21}^{-1}-K_{22}^{\dagger }K_{21}^{\dagger
-1}K_{21}^{-1}+K_{22}^{\dagger }\right) K_{21} \\
&=&K_{22}^{\dagger }\left( K_{22}^{\dagger }K_{21}^{\dagger
-1}K_{11}^{\dagger }K_{11}K_{21}^{-1}-K_{21}^{\dagger
-1}K_{21}^{-1}+I_{n_{2}}\right) K_{21} \\
&\overset{\ref{B2}}{=}&0;
\end{eqnarray*}

\begin{eqnarray*}
&&K_{11}^{\dagger }K_{12}+K_{21}^{\dagger }K_{22} \\
&=&K_{21}^{\dagger }\left( \left( K_{12}^{\dagger }K_{11}K_{21}^{-1}\right)
^{\dagger }+K_{22}\right)  \\
&\overset{\ref{B5}}{=}&K_{21}^{\dagger } 
\left( 
K_{22}^{\dagger
}K_{21}^{\dagger -1}K_{11}^{\dagger }K_{11}K_{21}^{-1}-K_{22}^{\dagger
}K_{21}^{\dagger -1}K_{21}^{-1}\right) ^{\dagger } \\
&&+ K_{21}^{\dagger } K_{22}\\
&=&K_{21}^{\dagger }\left( K_{21}^{\dagger -1}K_{11}^{\dagger
}K_{11}K_{21}^{-1}-K_{21}^{\dagger -1}K_{21}^{-1}+I_{n_{2}}\right) K_{22} \\
&\overset{\ref{B2}}{=}&0;
\end{eqnarray*}
and 
\begin{eqnarray*}
&&K_{11}^{\dagger }K_{11}+K_{21}^{\dagger }K_{21} \\
&\overset{\ref{B2}}{=}&K_{11}^{\dagger }K_{11}+K_{21}^{\dagger }\left(
K_{21}^{\dagger -1}K_{21}^{-1}-K_{21}^{\dagger -1}K_{11}^{\dagger
}K_{11}K_{21}^{-1}\right) K_{21} \\
&=&I_{n_{2}},
\end{eqnarray*}
whence $K^{\dagger }K=I_{n}$ as required.


\begin{thebibliography}{9}
\bibitem{minimal} The requirement of minimality means that there are no uncontrollable or unobservable modes of the system: this is a
standard concept in linear control theory which may be established in the Kalman state-based theory, see \cite{Kimura}, \cite{Paganini}.
\end{thebibliography}
\end{document}